\documentclass[transaction]{IEEEtran}
\usepackage{graphicx}
\usepackage{epstopdf}
\usepackage[latin1]{inputenc}
\usepackage{amsmath,amssymb,amscd,latexsym,dsfont}
\usepackage[english]{babel}
\usepackage{tabularx,cite}
\usepackage{graphicx}
\usepackage{algpseudocode}
\usepackage{algorithm}
\usepackage{algorithmicx}
\usepackage{float}
\usepackage{multicol}
\usepackage{psfrag}
\usepackage{comment,psfrag,subfigure,enumerate}
\usepackage{color}
\usepackage{amsthm}

\ifCLASSINFOpdf
\else
\fi
\begin{document}
%
\title{On Noisy ARQ in Block-Fading Channels}
\author{\IEEEauthorblockN{Behrooz Makki, Alexandre Graell i Amat, \emph{Senior Member, IEEE} and Thomas Eriksson}\\
\thanks{The authors are with Department of Signals and Systems,
Chalmers University of Technology, Gothenburg, Sweden, Email: \{behrooz.makki, alexandre.graell, thomase\}@chalmers.se}
\thanks{Alexandre Graell i Amat was supported by the Swedish Agency for Innovation Systems (VINNOVA) under the P36604-1 MAGIC project.}
}

%
\maketitle
\vspace{-0mm}
\begin{abstract}
Assuming noisy feedback channels, this paper investigates the data transmission efficiency and robustness of different automatic repeat request (ARQ) schemes using adaptive power allocation. Considering different block-fading channel assumptions, the long-term throughput, the delay-limited throughput, the outage probability and the feedback load of different ARQ protocols are studied. A closed-form expression for the power-limited throughput optimization problem is obtained which is valid for different ARQ protocols and feedback channel conditions. Furthermore, the paper presents numerical investigations on the robustness of different ARQ protocols to feedback errors. It is shown that many analytical assertions about the ARQ protocols are valid both when the channel remains fixed during all retransmission rounds and when it changes in each round (in)dependently. As demonstrated, optimal power allocation is crucial for the performance of noisy ARQ schemes when the goal is to minimize the outage probability.
\end{abstract}
%
\IEEEpeerreviewmaketitle
\vspace{0mm}
%
%
\section{Introduction}
\vspace{-0mm}

Automatic repeat request (ARQ) is a well-established approach aiming towards high throughput reliable wireless communication \cite{tuninetti2011,5439316,5753988,ARQ20111,throughputdef,1661837,4533229,4595018,5336856,Tcomkhodemun,1200407,wirelesskhodemun,Arulselvan,1379007,4356994,4200959,5771499,ARQGlarsson}. Utilizing both forward error correction and error detection, ARQ techniques reduce the data outage probability and/or increase the throughput by retransmitting the data which has experienced \emph{bad} channel conditions. ARQ is a technique in the \emph{data link layer} already provided in many wireless protocols, e.g., IEEE 802.11n \cite{ieeestandard1} and IEEE 802.16e \cite{ieeestandard2}. Hence, it needs no additional design which introduces it as a cost- and complexity-efficient approach.


In wireless networks, the feedback signals reach the transmitter through a communication link experiencing different levels of noise and fading. Hence, it is probable to receive erroneous signals at the transmitter which, if not handled suitably, can degrade the system performance severely and make it even worse than an \emph{open-loop} system \cite{refdarsad4,noisyARQ,noisyARQ2,05672052,00554284,00725313,05426078,01599696,05724298}.
This is because, due to receivers limited power and the users interference constraints, the ARQ bits may be fed back at low powers and, consequently, may be received by the transmitters unreliably. Therefore, it is interesting to study the channel performance under noisy feedback conditions.

From another perspective, wireless systems are normally power constrained. Therefore, with limited power resources, optimal power allocation in the ARQ retransmission rounds is a key point for increasing the system data transmission efficiency \cite{wirelesskhodemun,4200959,1379007,4356994,Arulselvan,1200407,5771499,ARQGlarsson}. In the concept of \emph{green} communication, adaptive power allocation is crucial, as the energy consumption of the wireless network is expected to increase by $16-20\%$ every year and contributes about $2\%$ of the global $CO_2$ emissions \cite{Gartner}. Specifically, with noisy feedback channels optimal power allocation becomes even more effective, as a large part of the resources may be lost due to errors in the feedback decoding. Furthermore, as discussed in the paper, using optimal power allocation within the retransmission rounds not only increases the data transmission reliability but also reduces the expected delay for data transmission, leading to higher throughput.

\emph{Literature review:} The ARQ-based papers related to our work can be divided into two categories. The first group is the works that have studied the optimal power allocation, between the retransmission rounds, in noise-free feedback condition. Here, the goal of power allocation is to minimize the required number of retransmission rounds \cite{1200407},
bit error rate \cite{1379007}, average overflow rate \cite{4356994} and the outage probability \cite{4200959,5771499,ARQGlarsson}, or to maximize the throughput \cite{tuninetti2011,5336856,Arulselvan} and the outage-limited average rate \cite{Tcomkhodemun,wirelesskhodemun}. The results have been obtained in the presence of transmitter channel state information (CSI) \cite{tuninetti2011,Arulselvan,4356994,1379007}, in delay- and buffer-limited condition \cite{1379007,4356994} and when the channel changes in each retransmission round \cite{tuninetti2011,Arulselvan,4356994,4200959} or remains constant between the whole retransmissions \cite{1200407,5336856,Tcomkhodemun,1379007,5771499,ARQGlarsson}. Particularly,  \cite{tuninetti2011,5336856,Tcomkhodemun} have presented some theoretical comparisons between the repetition time diversity (RTD) and incremental redundancy (INR) hybrid ARQ protocols. However, the comparisons are given either under a specific continuous communication model assumption with uniform power allocation \cite{5336856,Tcomkhodemun} or for fixed-length INR ARQ in the presence of transmitter partial CSI \cite{tuninetti2011}. In all these works, the feedback signal is received error-free.

The second group, on the other hand, are the papers that have investigated the effect of feedback channel noise on the performance of the ARQ protocols, e.g., \cite{05672052,00554284,00725313,05426078,01599696,noisyARQ,noisyARQ2,05724298}. However, in none of these works adaptive power allocation has been considered, and the results have been presented for the case of uniform power allocation. Moreover, there is no general framework that unifies the analysis of noisy ARQ protocols from an information theoretic perspective.

\emph{Contributions:} This paper demonstrates a fairly general information-theoretic framework for studying the noisy ARQ protocols utilizing adaptive power allocation. We obtain the results under different fading channel assumptions.
The discussions that we present here have not been covered in the reviewed papers; A closed-form expression for the power-limited throughput optimization problem is presented which is valid for the considered ARQ protocols and any feedforward or feedback channel conditions. Then, the long-term throughput, the delay-limited throughput and the outage probability of different ARQ protocols are obtained and compared in noisy feedback conditions. Also, both fixed- and variable-length coding hybrid ARQ (HARQ) approaches are investigated. Finally, the robustness of the protocols to feedback channel errors is compared numerically.

The data transmission efficiency of HARQ protocols are normally studied under two different assumptions where the channel is supposed to be fixed within all retransmission rounds, e.g., \cite{5336856,Tcomkhodemun,wirelesskhodemun,1200407}, or changing in each round (in)dependently, e.g., \cite{tuninetti2011,ARQ20111,4200959,1379007,4356994}. In other words, \cite{5336856,Tcomkhodemun,wirelesskhodemun,1200407} assume the blocks to be so long (or the codewords so short) that all retransmissions experience the same fading condition. On the other hand, in \cite{tuninetti2011,ARQ20111,4200959,1379007,4356994} the length of the codewords is supposed to be the same as the fading block length. In this paper, the developed framework is used to study both cases in detail. Specifically, it is shown that the analytical assertions of the paper are valid in both scenarios.
This point provides an appropriate connection between the papers considering one of these assumptions.

The main results of the paper are summarized as follows. Compared to the fixed-length coding scheme, the throughput and the robustness of the INR protocol increases when variable-length coding is implemented.
In terms of throughput and with different power allocation schemes, the ARQ protocols are observed to have low sensitivity to small feedback bit error probabilities. However, the sensitivity of the throughput to feedback channel noise and the effect of optimal power allocation increases with the number of retransmissions. Also, depending on the fading distribution, optimal power allocation can improve the relative throughput, defined as the normalized difference between the throughput achieved by optimal and uniform power allocation, and increases the robustness of the ARQ protocols with respect to the open-loop communication setup. Furthermore, optimal power allocation plays an important role on the performance of noisy and noise-free ARQ schemes when the goal is to minimize the outage probability.  Particularly, optimal power allocation results in considerable outage probability reduction although, compared to noise-free feedback conditions, the reduction is less pronounced when the feedback signal is unreliable.

With a noisy feedback channel, new analytical comparisons between the INR and RTD protocols are presented which show the superiority of the INR approach in different points of view. However, it is proved that the performance of these protocols converge at low signal-to-noise ratios (SNRs). Also, compared to the INR, the RTD is observed to be more robust to feedback channel noise, in the sense that the performance loss in the RTD is less than in the INR. Finally,
the difference between the optimal powers of the (re)transmissions increases when the feedback channel noise increases or the forward channel variability decreases.

\emph{Notation.} The following notation is used throughout the paper:
\begin{itemize}
  \item A packet is defined as the transmission of a codeword along with all its possible retransmission rounds. Also, we consider a maximum of $M$ retransmission rounds, i.e., each codeword is (re)transmitted a maximum of $M+1$ times.
  \item  $l_m$ (in channel uses) is the length of the codeword (re)transmitted in the $m$-th round.
  \item $R^{(m)}$ denotes the \emph{equivalent} transmission rate at the end of the $m$-th (re)transmission round. Thus, denoting the number of information nats considered for a packet by $Q$, we have $R^{(m)}=\frac{Q}{l^{(m)}},l^{(m)}=\sum_{n=1}^{m}{l_n}$.
  \item $\Pr(\text{Outage})$ represents the outage probability, i.e., the probability of the event that the data can not be decoded by the receiver when the data (re)transmission is stopped. Therefore, the expected number of nats that is successfully decoded in each packet period is $\bar Q=Q(1-\Pr(\text{Outage}))$.
  \item $P_m$ is the transmission power used per channel use in the $m$-th (re)transmission round. Consequently, $\xi_m=P_m l_m$ is the energy consumed in the $m$-th round.
  \item $\Pr ({A_m})$ represents the probability of the event that data (re)transmission is stopped at the end of the $m$-th round. In this case, due to possible errors in the feedback bits, the data sent by the transmitter might have been decoded or not by the receiver in the $n$-th, $n\le m$, (re)transmission rounds. Also, as a maximum of $M+1$ (re)transmission rounds are permitted, $\sum_{m = 1}^{M + 1} {\Pr ( {A_m}) }  = 1$.
  \item ${R^{(0)}} \buildrel\textstyle.\over= \infty$ and $A_0=\emptyset$ denotes the empty set.
  \item $\Pr({S_m})$ is the probability that while the data has been \emph{successfully} decoded in one of the time slots $n=1,\ldots,m$, the data transmission is stopped at the $m$-th (re)transmission round. In contrast to $A_m$, the event $S_m$ does not include the case where data (re)transmission stops while the codeword has not been decoded by the receiver yet.
\end{itemize}
\vspace{-0mm}
\section{System model}
\vspace{-0mm}
Consider a block-fading channel where the fading coefficient remains constant for a duration
of $L_\text{c}$ channel uses, generally determined by the channel coherence time, and changes independently from one block to another. In the $m$-th (re)transmission round of a packet, the received signal is obtained by
\begin{align}
  Y_m[i]=\sqrt{P_m}hX_m[i]+Z_m[i],i=1,\ldots,l_m.
\end{align}
Here, $X_m[i],i=1,\ldots,l_m,\,\frac{1}{l_m}\sum_{i=1}^{l_m}{|X_m[i]|^2}=1,$ is the power-limited transmission codeword, $h$ is the channel coefficient, $Z_m[i] \sim \mathcal{CN}(0,1)$ denotes an independent and identically distributed (i.i.d.) complex Gaussian noise and $P_m$ is the transmission power in the $m$-th round that, because the noise variance is set to 1, represents the transmission SNR as well (in dB, the SNR is given by $10\log_{10}(P_m)$). Also, we define $g=|h|^2$ as the \emph{channel gain} random variable which follows the fading probability density function (pdf) ${f_G}(g)$.

%

The receiver is assumed to have perfect instantaneous CSI, which is an acceptable assumption under block-fading \cite{tuninetti2011,5439316,5753988,ARQ20111,1661837,4533229,4595018,Tcomkhodemun,5336856,throughputdef,33}. On the other hand, there is no \emph{instantaneous} CSI available at the transmitter, except the ARQ feedback bits\footnote{Further discussion about the transmitter CSI is given in Section III.}.
The feedback channel is supposed to be noisy with bit error probability $p_\text{b}$. Moreover, as each transmission experiences an AWGN channel, all results are restricted to Gaussian input distributions. Finally, the results are presented in natural logarithm basis, unless otherwise stated, and the throughput is given in nats-per-channel-use (npcu). Two different assumptions are considered for the length of the blocks throughout the paper:
\begin{itemize}
  \item[1)] Long-$L_\text{c}$ scenario: In this case, the length of the blocks, $L_\text{c}$, is assumed to be so long that all retransmission rounds occur in a single fading block. That is, the channel is supposed to remain fixed during a packet transmission period and change independently from one packet to another. This is an appropriate  model for networks with stationary or slow-moving users \cite{wirelesskhodemun,Tcomkhodemun,5336856,1200407}.
  \item[2)] Short-$L_\text{c}$ scenario: In Section VI, the codewords lengths are considered to be the same as the fading block length $L_\text{c}$ such that the channel changes in each retransmission round. The results of this part are useful for modeling users with medium/fast speeds \cite{tuninetti2011,ARQ20111,4200959,1379007,4356994}.
\end{itemize}

\vspace{-0mm}
\section{Long-term throughput analysis}
Here, the long-term throughput \cite{throughputdef} and the average transmission power \cite{excellentref} are respectively defined as
\vspace{-0mm}
\begin{align}
\eta_\text{LT}  \buildrel\textstyle.\over= \frac{{\bar Q}}{{\bar \tau }},
\end{align}
\vspace{-0mm}
and
\vspace{-0mm}
\begin{align}
\phi  \buildrel\textstyle.\over= \frac{{\bar \xi }}{{\bar \tau }}
\end{align}
where ${\bar Q}$, ${\bar \tau }$ and ${\bar \xi }$ denote the expected value of the successfully decoded information nats, the expected number of channel uses and the expected energy consumed within a packet transmission period, respectively.

In the following, first a closed-form expression for the power-limited long-term throughput optimization problem is derived. The expression is valid for any feedback channel conditions and all HARQ protocols that are considered in this work. Later, the long-term throughput is analyzed in more detail for the basic ARQ, RTD and INR HARQ.

\emph{Lemma 1:} Independent of the forward or feedback channel condition, the power-limited long-term throughput optimization problem of an ARQ scheme with a maximum of $M$ retransmission rounds and power constraint $P$ can be expressed as
\vspace{-0mm}
\begin{align}
 &\mathop {\max }\limits_{\forall {P_m},{R^{(m)}}} \frac{{1-\Pr ( \text{Outage}) }}{{\sum_{m = 1}^{M + 1} {\frac{{\Pr ( {A_m}) }}{{{R^{(m)}}}}} }} \\&
 \,\,\,\,\,\text{s.t.}\,\,\frac{{\sum_{m = 1}^{M + 1} {{P_m}\left(\frac{1}{{{R^{(m)}}}} - \frac{1}{{{R^{(m - 1)}}}}\right)\left(1 - \sum_{n = 1}^{m - 1} {\Pr ( {A_n}) } \right)} }}{{\sum_{m = 1}^{M + 1} {\frac{{\Pr ( {A_m}) }}{{{R^{(m)}}}}} }}\le P
\end{align}
where (4) and (5) represent the long-term throughput and the average transmission power, respectively.

\begin{proof} To calculate the long-term throughput, assume that $Q$ information nats are transmitted in each packet transmission. If the data is successfully decoded at any (re)transmission round, all the $Q$ nats are received by the receiver. Hence, as stated before, we have
\vspace{-0mm}
\begin{align}
{\bar Q}=Q \left(1-\Pr(\text{Outage})\right).\vspace{-0mm}
\end{align}
On the other hand, if the data (re)transmission is, either successfully or not, stopped at the end of the $m$-th (re)transmission round, the total number of channel uses is $\sum_{n = 1}^m {{l_n}}$. Therefore, the expected number of channel uses within a packet transmission period is
\vspace{-0mm}
\begin{align}
\bar \tau  = \sum_{m = 1}^{M + 1} {\left(\sum_{n = 1}^m {{l_n}} \right)\Pr ( {A_m}) }.
\end{align}
In this way, from (2), (6), (7) and as the equivalent transmission rate at the end of the $m$-th (re)transmission round is ${R^{(m)}} = \frac{Q}{{\sum_{n = 1}^m {{l_n}} }}$, the long-term throughput is found as stated in (4).

Provided that the data (re)transmission ends at the $m$-th round, the total consumed energy is ${\xi ^{(m)}} = \sum_{n = 1}^m {{P_n}{l_n}}$. Therefore, the expected energy consumed within a packet transmission period is obtained by
\begin{align}
\begin{array}{l}
\bar \xi  = \sum_{m = 1}^{M + 1} {\left(\sum_{n = 1}^m {{P_n}{l_n}} \right)\Pr ( {A_m}) } \\\,\,\,\,\mathop  = \limits^{(a)} Q\sum_{m = 1}^{M + 1} {\left(\sum_{n = 1}^m {{P_n}\left(\frac{1}{{{R^{(n)}}}} - \frac{1}{{{R^{(n - 1)}}}}\right)} \right)\Pr ( {A_m}) }  \\\,\,\,\,=Q \sum_{m = 1}^{M + 1} {{P_m}\left(\frac{1}{{{R^{(m)}}}} - \frac{1}{{{R^{(m - 1)}}}}\right)\left(1 - \sum_{n = 1}^{m - 1} {\Pr ( {A_n}) } \right)}\\
\end{array}
\end{align}
where $(a)$ is due to the fact that ${l_m} = \frac{Q}{{{R^{(m)}}}} - \frac{Q}{{{R^{(m - 1)}}}}$. Finally, from (3), (7), (8) and ${R^{(m)}} = \frac{Q}{{\sum_{n = 1}^m {{l_n}} }}$, the average transmission power is rephrased as (5). Therefore, the power-limited long-term throughput optimization problem can be expressed as stated in (4)-(5).
\end{proof}

Notice that with uniform power allocation, the power constraint (5) simplifies to $P_m=P'\le P$. Then, as the achievable rate of the AWGN channel is an increasing function of the transmission power \cite{MMSE1}, maximizing the achievable rate implies $P_m=P, \forall m$. Finally, to find
the throughput of different ARQ protocols, it is only required to determine their corresponding probability terms in (4) and (5).

\emph{Corollary 1:} For fixed-length coding schemes, the maximum power-limited long-term throughput is obtained by
\vspace{-0mm}
\begin{align}
 &\mathop {\max }\limits_{\forall {P_m},{R}} \frac{{R(1-\Pr ( \text{Outage})) }}{{\sum_{m = 1}^{M + 1} {m\Pr ( {A_m}) } }} \\
 &\,\,\,\text{s.t.}\,\,\,\frac{{\sum_{m = 1}^{M + 1} {{P_m}\left(1 - \sum_{n = 1}^{m - 1} {\Pr ( {A_n}) } \right)} }}{{\sum_{m = 1}^{M + 1} {m\Pr ( {A_m}) } }}\le P
\end{align}
where $R=\frac{Q}{L}$ is the initial codeword rate and $L$ is the length of the codewords.
\begin{proof}
Using $l_m=L \,\forall m$, we have $R^{(m)}=\frac{Q}{mL}=\frac{R}{m}$ which rephrases (4) and (5) as in (9) and (10), respectively.
\end{proof}


In the sequel, the general equations (4) and (5) are specialized for the RTD and the INR HARQ protocols under a noisy feedback channel assumption. Performance analysis for the basic ARQ schemes can be found in the appendix\footnote{Throughout the paper, whenever required, the results are particularized for the long- and short-$L_\text{c}$ scenarios. If not mentioned, the discussions are valid for both cases.}.

\vspace{-0mm}
\subsection{RTD protocol in the long-$L_\text{c}$ scenario}
Utilizing the RTD (also called \emph{Chase combining}) HARQ, the same codeword is (re)transmitted in each (re)transmission round and the receiver performs maximum ratio combining of the received signals. Therefore, assuming the long-$L_\text{c}$ scenario, the received SNR after $m$ data (re)transmission rounds increases to $g\sum_{n=1}^{m} {P_n}$ and the equivalent data rate reduces to $R^{(m)}=\frac{R}{m}$. The data is correctly decoded at the end of the $m$-th round (and not before) if 1) all previous feedback bits have been correctly decoded by the transmitter (with probability $(1-p_\text{b})^{m-1}$), 2) the receiver has not decoded the data before, i.e., $\log (1 + g\sum_{j = 1}^{n} {{P_j}} ) < R,\, \forall n<m$, and 3) (re)transmitting the data in the $m$-th slot, the receiver can decode the codeword, i.e., $\log (1 + g\sum_{n = 1}^m {{P_n}} ) \ge R$. Hence, the data outage probability is found as
\vspace{-0mm}
\begin{align}
\Pr ( \text{Outage})^{\text{RTD}}  = 1-\sum_{m = 1}^{M + 1} {{{(1 - {p_\text{b}})}^{m - 1}}\Pr {{( m) }^{\text{RTD}}}}
\end{align}\vspace{-0mm}
where
\vspace{-0mm}
\begin{align}
\Pr {( m) ^{\text{RTD}}} &= \Pr ( \log (1 + g\sum_{n = 1}^{m - 1} {{P_n}} ) < R \le \log (1 + g\sum_{n = 1}^m {{P_n}} ))\nonumber\\&= {F_G}(\frac{{{e^R} - 1}}{{\sum_{n = 1}^{m - 1} {{P_n}} }}) - {F_G}(\frac{{{e^R} - 1}}{{\sum_{n = 1}^m {{P_n}} }})
\end{align}
is the probability that, assuming a noise-free feedback channel, the data is decoded at the end of the $m$-th time slot while it was not decodable before. Also, $F_G$ represents the channel gain cumulative distribution function (cdf). Note that in (12) we have used the fact that with an equivalent SNR $x$ the maximum decodable transmission rate is $\frac{1}{m} \log(1+x)$ if a codeword is repeated $m$ times.

On the other hand, with some manipulations, the probability that, either successfully or not, the data transmission stops at the $m$-th (re)transmission round is
\begin{align}
\begin{array}{l}
 {\Pr(A_m)^\text{RTD}} = \sum_{n = 1}^m {\Pr ( n)^\text{RTD} } {(1 - {p_{\text{b}}})^n}p_{\text{b}}^{m - n}
  \\\,\,\,\,\,+ \left( {1 - \Pr ( 1,2, \ldots ,m)^\text{RTD} } \right){(1 - {p_{\text{b}}})^{m - 1}}{p_{\text{b}}} \\\,\,\,\,\,
  = \sum_{n = 1}^m {{{(1 - {p_{\text{b}}})}^n}p_{\text{b}}^{m - n}\left( {{F_G}(\frac{{{e^{{R}}} - 1}}{{\sum_{k = 1}^{n - 1} {{P_k}} }}) - {F_G}(\frac{{{e^{{R}}} - 1}}{{\sum_{k = 1}^n {{P_k}} }})} \right)}
 \\\,\,\,\,\, + {F_G}(\frac{{{e^{{R}}} - 1}}{{\sum_{k = 1}^m {{P_k}} }}){(1 - {p_{\text{b}}})^{m - 1}}{p_{\text{b}}},\,m = 1, \ldots ,M, \\
 {\Pr(A_{M + 1})^\text{RTD}} = \sum_{n = 1}^{M + 1} {\Pr ( n)^\text{RTD} } {(1 - {p_{\text{b}}})^{n - 1}}p_{\text{b}}^{M + 1 - n} \\\,\,\,\,\,+ (1 - \Pr ( 1,2, \ldots ,M + 1)^\text{RTD} ){(1 - {p_{\text{b}}})^M} \\
 \end{array}
\end{align}
where
\vspace{-0mm}
\begin{align}
\Pr ( 1,2,\ldots,m)^{\text{RTD}}  &= \Pr \left( R \le \log (1 + g\sum_{k = 1}^m {{P_k}} )\right)  \nonumber\\&= 1 - {F_G}(\frac{{{e^{{R}}} - 1}}{{\sum_{k = 1}^m {{P_k}} }})
\end{align}
is the probability that the data is decodable in one of the first $m$ (re)transmission rounds if all feedback bits are correctly decoded. Finally, using (11) and (13) in (9) and (10), the long-term throughput and the average transmission power for the RTD protocol are obtained. Note that setting $p_\text{b}=0$, i.e., noise-free feedback channel, we have $\Pr(A_m)=\Pr(m),\, m=1,\ldots,M,$  and $\Pr(A_{M+1})=1-\sum_{m=1}^{M}{\Pr(m)}$. Also, with uniform power allocation, i.e., $P_m=P\,\forall m$, (12) and (14) respectively change to
\vspace{-0mm}
\begin{align}
\Pr {( m) ^{\text{RTD}}} = {F_G}(\frac{{{e^R} - 1}}{{(m - 1)P}}) - {F_G}(\frac{{{e^R} - 1}}{{mP}}),
\end{align}
\vspace{-0mm}
and
\vspace{-0mm}
\begin{align}
\Pr {( 1,2, \ldots ,m) ^{\text{RTD}}} = 1 - {F_G}(\frac{{{e^R} - 1}}{{mP}}).
\end{align}
\vspace{-0mm}
\subsection{INR protocol in the long-$L_\text{c}$ scenario}
INR is a well-known HARQ scheme where, at each retransmission round, new redundancy bits are sent by the transmitter and the receiver combines them. The probabilities  $\Pr(\text{Outage})^{\text{INR}}$ and ${\Pr(A_m)^\text{INR}},\,m=1,\ldots,M+1,$ are determined with the same procedure as for the RTD protocol while, using the results of \cite{excellentref} and \cite[chapter 15]{4444444444}, the terms ${\Pr(m)^\text{INR}}$ and $\Pr (1,2,\ldots,m)^{\text{INR}}$ are respectively obtained by the time division multiple access (TDMA)-type equations
\vspace{-0mm}
\begin{align}
\begin{array}{l}
 \Pr {( m) ^{\text{INR}}} = \Pr \bigg( \sum_{n = 1}^{m - 1} {\frac{{{l_n}}}{{\sum_{j = 1}^{m - 1} {{l_j}} }}\log (1 + g{P_n})}  < {R^{(m - 1)}}\, \&\\\,\,\,\,\,\,\,\,\,\,\,\,\,\,\,\,\,\,\,\,\,\,\,\,\,\,\,\,\,\,\,\,\,\,\,\,\,\,\,\,\,\,\,\,\,\,\,\,\,\,\,\,\,\,\, \sum_{n = 1}^m {\frac{{{l_n}}}{{\sum_{j = 1}^m {{l_j}} }}\log (1 + g{P_n})}  \ge {R^{(m)}}\,\,\bigg)  \\
  \,\,\,\,\,\,\,\,\,\,\,\,\,\,\,\,\,\,\,\,\,\,\,\,\,\,= \Pr \bigg( \sum_{n = 1}^{m - 1} {(\frac{1}{{{R^{(n)}}}} - \frac{1}{{{R^{(n - 1)}}}})\log (1 + g{P_n})}  < 1\\\,\,\,\,\,\,\,\,\,\,\,\,\,\,\,\,\,\,\,\,\,\,\,\,\,\,\,\,\,\,\,\,\,\,\,\,\,\,\,\,\,\,\,\,\,\,\,\,\,\,\,\,  \le \sum_{n = 1}^m {(\frac{1}{{{R^{(n)}}}} - \frac{1}{{{R^{(n - 1)}}}})\log (1 + g{P_n})} \bigg)  \\
 \end{array}
\end{align}
and
\vspace{-0mm}
\begin{align}
&\Pr {( 1,2, \ldots ,m) ^{\text{INR}}} = \Pr ( \sum_{n = 1}^m {\frac{{{l_n}}}{{\sum_{j = 1}^m {{l_j}} }}\log (1 + g{P_n})}  \ge {R^{(m) }}) \nonumber\\& = \Pr ( \sum_{n = 1}^m {(\frac{1}{{{R^{(n)}}}} - \frac{1}{{{R^{(n - 1)}}}})\log (1 + g{P_n})}  \ge 1),
\end{align}
if the channel does not change in the (re)transmission rounds. Here, e.g., (17), follows from the fact that using $m$ different codewords of length $l_n$ and power $P_n,\,n=1,\ldots,m$, the maximum decodable information rate is $\sum_{n = 1}^m {\frac{{{l_n}}}{{\sum_{j = 1}^m {{l_j}} }}\log (1 + g{P_n})}$. Then, with uniform power allocation (17) and (18) are respectively rephrased as
\vspace{-0mm}
\begin{align}
\Pr {( m) ^{\text{INR}}} = {F_G}(\frac{{{e^{{R^{(m - 1)}}}} - 1}}{P}) - {F_G}(\frac{{{e^{{R^{(m)}}}} - 1}}{P})
\end{align}
and
\vspace{-0mm}
\begin{align}
\Pr {( 1,2, \ldots ,m) ^{\text{INR}}} = 1 - {F_G}(\frac{{{e^{{R^{(m)}}}} - 1}}{P}).
\end{align}
Also, considering fixed-length coding, i.e., $l_m=L\, \forall m$ in (17) and (18), we have $R^{(m)}=\frac{R}{m}$,
\vspace{-0mm}
\begin{align}
\Pr {( m) ^{\text{INR}}} = \Pr \left( \sum_{n = 1}^{m - 1} {\log (1 + g{P_n})}  < R\, \le \sum_{n = 1}^m {\log (1 + g{P_n})} \right)
\end{align}
and
\vspace{-0mm}
\begin{align}
 \Pr {( 1,2, \ldots ,m) ^{\text{INR}}} = \Pr \left( \sum_{n = 1}^m {\log (1 + g{P_n})}  \ge R\right).
\end{align}
Using (17)-(22) in (4) and (5), we can obtain the power-limited long-term throughput of the INR approach. Also, the probability terms of (17)-(18) are obtained via
\begin{align}
&\Pr\left(\sum_{n=1}^{m}{(\frac{1}{R^{(n)}}-\frac{1}{R^{(n-1)}})\log(1+gP_n)}\le 1\right)=F_G(\Delta_m),
\nonumber\\&\Delta_m=\mathop {\arg }\limits_g \{\sum_{n=1}^{m}{(\frac{1}{R^{(n)}}-\frac{1}{R^{(n-1)}})\log(1+gP_n)}= 1)\}.\nonumber
\end{align}
Considering different values of $m$, there is no general closed-form solution for $\Delta_m$. Therefore, depending on the fading distribution and the number of retransmissions, $\Delta_m$ may need to be numerically calculated. However, as $R^{(n)}<R^{(n-1)},\forall n,$ the function $\Theta_m(g)=\sum_{n=1}^{m}{(\frac{1}{R^{(n)}}-\frac{1}{R^{(n-1)}})\log(1+gP_n)}$ is an increasing function of $g$ and, therefore, $\Delta_m$ is unique for a given set of $\{P_n, R^{(n)}, n=1,\ldots,m\}$.

\subsection{Discussions}
We close this section with discussions about the optimization problem of (4)-(5) and (9)-(10) and some practical implementation issues of the proposed scheme.

Using the same arguments as in \cite{5336856,Tcomkhodemun,wirelesskhodemun,5771499}, it can be showed that both the power-limited throughput maximization and the outage probability minimization of HARQ protocols are nonconvex optimization problems, even if the feedback channel is noise-free. Therefore, the problems should be solved via iterative optimization algorithms. In our setup, the number of optimization parameters is low enough to use exhaustive search, which is what we have used for our simulations. Also, for faster convergence, we have repeated the simulations by using the iterative algorithm of \cite{Tcomkhodemun}, and by using ``fminsearch'' and ``fmincon'' functions of MATLAB. The results have been obtained for different initial settings and we have tested the fmincon function for ``\emph{interior-point},'' ``\emph{active-set}'' and ``\emph{trust-region-reflective}'' options of the optimization algorithm. In all cases, the results are the same with high accuracy, which is an indication of a reliable result. In our experiments, the exhaustive search and the fmincon-based codes are, respectively, the slowest and the fastest schemes, compared to fminsearch and the iterative algorithm of \cite{Tcomkhodemun}. However, as the parameters of, e.g., (4)-(5), are determined off-line, the complexity is not as important as in online applications.


In practice, the suitable transmission parameters can be determined in two ways. In the first method, the system performance is evaluated off-line for different rates/powers, and the appropriate parameters are collected in a table which is used during data transmission. In this case, which is the same as in adaptive modulation and coding (AMC) protocols \cite{4381368}, there is no need to know the channel cdf (in general, the only parameters that we need to know are the rates and powers $R^{(m)}$ and $P^{(m)},\forall m,$ and not the channel cdf.). In the second method, however, the gain cdf and an optimization algorithm are utilized by the transmitter for parameter setting.  This is a suitable method for the scenario where the channel follows a specific cdf pattern and only the long-run statistics (e.g., the gain mean and variance) change after several packet periods. Thus, the amount of feedback required for long-run adaptation of the statistics is negligible, compared to the ARQ feedback bits, and the gain cdf can be assumed to be known by the transmitter, in harmony with \cite{throughputdef,1661837,4533229,4595018,5336856,Tcomkhodemun,1200407,wirelesskhodemun}.

Finally, the power allocation between the (re)transmissions is carried out through an adaptive power controller at the transmitter, the same as in \cite{1200407,1379007,4200959,5771499,ARQGlarsson,Tcomkhodemun,wirelesskhodemun,5336856,Arulselvan,4356994} that deal with power allocation for cases with a noise-free feedback channel. Here, the only difference with \cite{1200407,1379007,4200959,5771499,ARQGlarsson,Tcomkhodemun,wirelesskhodemun,5336856,Arulselvan,4356994} is in the values of the (re)transmission rates/powers which are selected such that the system performance is optimized when there is an error probability in decoding the feedback bits.

\vspace{-0mm}
\section{On the performance of RTD and INR protocols in the long-$L_\text{c}$ scenario}
Considering a noise-free feedback channel, e.g., \cite{tuninetti2011,5336856,Tcomkhodemun} have presented comparisons between the RTD and INR protocols. Assuming the long-$L_\text{c}$ scenario, this section presents new analytical results for the HARQ protocols in the case of a noisy feedback channel, which are the extensions of the results in \cite{tuninetti2011,5336856,Tcomkhodemun}. The theorems and analysis of this section are required for Section VI, where the equivalency of the short- and long-$L_\text{c}$ scenarios is demonstrated.



\vspace{-0mm}
\subsection{Feedback load and the expected number of (re)transmission rounds}
One of the most important aspects that quantifies the performance of limited feedback schemes such as ARQ protocols is the feedback load defined as the expected number of feedback bits transmitted in a packet period. On the other hand, the expected number of (re)transmission rounds is another metric demonstrating the average number of handshakings between the transmitter and the receiver within a packet transmission period. The following theorem compares the RTD and the INR protocols in terms of the expected number of (re)transmission rounds and the feedback load.

\emph{Remark 1:} With fixed-length coding, the expected number of (re)transmission rounds is the expected number of channel uses or the expected delay for a packet transmission scaled by a constant.

\emph{Theorem 1: } Assume uniform power allocation. Then, with the same feedback load (or the same expected number of (re)transmission rounds) higher throughput is achieved by the INR protocol when compared with the RTD.
\begin{proof}
Please see the appendix.
\end{proof}

Note that in Theorem 1 the rates ${R^{(m)}} = \log (1 + \frac{{{e^{\hat R}} - 1}}{m}) \ge \frac{\hat R}{m}$ are achievable by the INR protocol using variable-length coding.
\vspace{-0mm}
\subsection{Fixed-length coding}
In comparison to the RTD protocol, the INR HARQ is a complex scheme as not only new parity bits are sent in each (re)transmission round but also the length of the codewords may be different in the retransmission rounds. In order to reduce the implementation complexity of the INR protocol, fixed-length coding is a sub-optimal scheme considered in the literature \cite{tuninetti2011,ARQGlarsson,throughputdef}.

In the following, fixed-length coding is considered to show some of the properties of the INR. Let us first review a simple point which is used throughout the paper repeatedly;
%
Define the function $J(x) = \log (1 + ax) + \log (1 + by) - \log (1 + ax + by)$. Then, as $J(0)=0$ and $\frac{{dJ}}{{dx}} \ge 0,\,\forall x,y,a,b \ge 0$, it is concluded that
\vspace{-0mm}
\begin{align}
\log (1 + ax) + \log (1 + by) \ge \log (1 + ax + by),\,\forall x,y,a,b \ge 0.
\end{align}

We use (23) to show the superiority of INR over RTD in noisy feedback conditions.

\emph{Theorem 2: } With adaptive power allocation and for any feedback channel bit error probability, higher power-limited throughput is obtained by the INR-based HARQ when compared to the RTD-based scheme.
\begin{proof}
Please see the appendix.
\end{proof}

\emph{Corollary 2:} With adaptive power allocation, lower outage probability is obtained by the INR-based HARQ, when compared with the RTD approach.
\begin{proof}
As the outage probability in both schemes is obtained by $\Pr ( \text{Outage})^{\text{H}}  = 1-\sum\limits_{m = 1}^{M + 1} {{{(1 - {p_\text{b}})}^{m - 1}}\Pr {{( m) }^{\text{H}}}},$ $\text{H}=\{\text{RTD},\text{INR}\}$, the same argument as in Theorem 2 can be used to prove the corollary. Note that, although Theorem 2 and Corollary 2 have been proved for fixed-length coding, they are valid for the variable-length INR scheme as well.
\end{proof}

\emph{Theorem 3:} Utilizing fixed-length coding and for any feedback channel conditions, the power-limited throughput and the outage probability of the INR and RTD protocols are the same for low SNRs.
\begin{proof}
As $\log (1 + x) \simeq x$ for small values of $x$, the probabilities (12), (14), (21) and (22) are changed to
\vspace{-0mm}
\begin{align}
\Pr {( m) ^{\text{RTD}}} = \Pr {( m) ^{\text{INR}}} \simeq \Pr ( g\sum_{n = 1}^{m - 1} {{P_n}}  < R\, \le g\sum_{n = 1}^m {{P_n}} )
\end{align}
and
\vspace{-0mm}
\begin{align}
\Pr {( 1,2, \ldots ,m) ^{\text{RTD}}} = \Pr {( 1,2, \ldots ,m) ^{\text{INR}}} \simeq \Pr ( g\sum_{n = 1}^m {{P_n}}  \ge R)
\end{align}
which, from (9)-(10), lead to the same throughput, outage probability and average power in both schemes.
\end{proof}

Finally, it is worth noting that, as uniform power allocation is a special case of adaptive power allocation, the results of the section are valid for the case of uniform power allocation as well.

\vspace{-0mm}
\section{Delay-limited throughput}
Along with the long-term throughput, the \emph{delay-limited throughput}, defined as the expectation of the achievable rate within a packet transmission period, is another metric which is sometimes used to characterize the system performance. Generally, the long-term throughput is useful when considering the steady-state behavior of several packet transmissions as time goes to infinity \cite{tuninetti2011,5336856,throughputdef}. On the other hand, the delay-limited throughput is more capable to track the short time variations \cite{4533229,4595018}.

Extending the results of \cite{4533229,4595018} to noisy feedback conditions, the delay-limited throughput of an HARQ-based system is obtained by
\vspace{-0mm}
\begin{align}
{\eta _{\text{DL}}} = \sum_{m = 1}^{M + 1} {{R^{(m)}}\Pr ( {S_m}) }
\end{align}
where $\Pr ( {S_m})$ is the probability that while the data has been decoded in one of the time slots $n=1,\ldots,m$, the data transmission is stopped at the $m$-th (re)transmission round. Note that the probability term $\Pr ( {S_m})$ contains the events that while the data has been decoded at time slot $n\le m$, due to wrong decoding of the feedback bits, data (re)transmission has continued until the $m$-th round. In this case, the equivalent achievable rate is $R^{(m)}=\frac{Q}{\sum_{n=1}^{n=m}{l_n}}$. In other words, (26) is the expectation of the achievable rate during a packet transmission period. Also, note that
\vspace{-0mm}
\begin{align}
\Pr ( {S_m})  \le \Pr ( {A_m}),
\end{align}
as ${S_m} \subseteq {A_m}$, and
\vspace{-0mm}
\begin{align}
\sum_{m = 1}^{M + 1} {\Pr ( {S_m}) }  = 1 - \Pr ( \text{Outage}).
\end{align}
%


\emph{Remark 2:} With an average power constraint, the maximum delay-limited throughput of an HARQ scheme is obtained by replacing (26) in (4).

\emph{Theorem 4:} Considering delay-limited throughput, the following assertions are valid:
\begin{itemize}
  \item[(I)] ${\eta _{\text{DL}}} \ge (1 - \Pr ( \text{Outage}) )\eta_\text{LT} $.
  \item[(II)] Theorems 1, 2 and 3 are valid for the delay-limited throughput as well.
  \item[(III)] With noise-free feedback channel and optimal power allocation, the delay-limited throughput for the RTD and INR protocols under long-$L_\text{c}$ scenario are respectively obtained by
      \vspace{-0mm}
      \begin{align}
      \eta _{\text{DL}}^{\text{RTD}} = \sum_{m = 1}^{M + 1} {\frac{R}{m}\left({F_G}(\frac{{{e^R} - 1}}{{\sum_{n = 1}^{m - 1} {{P_n}} }}) - {F_G}(\frac{{{e^R} - 1}}{{\sum_{n = 1}^m {{P_n}} }})\right)}
      \end{align}
      and
      \vspace{-0mm}
      \begin{align}
      \eta _{\text{DL}}^{\text{INR}} = \sum_{m = 1}^{M + 1} {{R^{(m)}}} \big({F_G}({\Delta _{m - 1}}) - {F_G}({\Delta _m})\big)
      \end{align}
      where ${\Delta _m} \buildrel\textstyle.\over= \mathop {\arg }\limits_g \{ \sum_{n = 1}^m {(\frac{1}{{{R^{(n)}}}} - \frac{1}{{{R^{(n - 1)}}}})\log (1 + g{P_n})}  = 1\} $.
\end{itemize}
\begin{proof}
Part (I) is proved based on the following inequalities:
\vspace{-0mm}
\begin{align}
\begin{array}{l}
 \frac{{{\eta _{\text{DL}}}}}{{(1 - \Pr ( \text{Outage}) )}} = \sum_{m = 1}^{M + 1} {(\frac{Q}{{\sum_{n = 1}^m {{l_n}} }})\frac{{\Pr ( {S_m}) }}{{(1 - \Pr ( \text{Outage}) )}}}
 \\ \,\,\,\,\,\,\,\,\,\,\,\,\,\,\,\,\,\,\,\,\,\,\,\,\,\,\,\,\,\,\,\,\,\mathop  \ge \limits^{(b)} \frac{{Q(1 - \Pr ( \text{Outage}) )}}{{\sum_{m = 1}^{M + 1} {(\sum_{n = 1}^m {{l_n}} )\Pr ( {S_m}) } }}\\ \,\,\,\,\,\,\,\,\,\,\,\,\,\,\,\,\,\,\,\,\,\,\,\,\,\,\,\,\,\,\,\,\,\mathop  \ge \limits^{(c)} \frac{{Q(1 - \Pr ( \text{Outage}) )}}{{\sum_{m = 1}^{M + 1} {(\sum_{n = 1}^m {{l_n}} )\Pr ( {A_m}) } }}\mathop  = \limits^{(d)} \frac{{1-\Pr ( \text{Outage}) }}{{\sum_{m = 1}^{M + 1} {\frac{{\Pr ( {A_m}) }}{{{R^{(m)}}}}} }} = \eta_\text{LT}  \\
 \end{array}
\end{align}
where $(b)$ is based on Jensen's inequality \cite{4444444444}, convexity of the function $f(x)=\frac{1}{x}$ and (28), $(c)$ comes from (27) and $(d)$ follows from $R^{(m)}=\frac{Q}{\sum_{n=1}^{n=m}{l_n}}$ and (4).

Part (II) is proved with the same procedure as for the Theorems 1, 2 and 3 while, with the same arguments as before, we have
\vspace{-0mm}
\begin{align}
\begin{array}{l}
 \Pr {( {S_m}) ^\text{H}} = \sum_{n = 1}^m {\Pr {{( n) }^\text{H}}p_\text{b}^{m - n}{{(1 - {p_\text{b}})}^n}} ,\,m = 1,\ldots,M \\
 \Pr {( {S_{M + 1}}) ^\text{H}} = \sum_{n = 1}^{M + 1} {\Pr {{( n) }^\text{H}}p_\text{b}^{M + 1 - n}{{(1 - {p_\text{b}})}^{n - 1}}},  \\
 \end{array}
\end{align}
where $\text{H}=\{\text{RTD},\text{INR}\}$. Finally, part (III) is obtained based on (12)-(14), (17), (18) and the fact that under noise-free feedback channel assumption we have $\Pr ( {S_m})  = \Pr ( {m}),\,m=1,\ldots,M+1$.
\end{proof}

\emph{Corollary 3:} The long-term throughput of an HARQ scheme is upper bounded by ${\eta _{\text{LT}}} \le \sum_{m = 1}^{M + 1} {{R^{(m)}}\Pr ( {A_m}) } $.

\begin{proof}
Similar to Theorem 4 part (I), the upper bound is obtained based on (4), the Jensen's inequality, convexity of the function $f(x)=\frac{1}{x}$, $\sum_{m = 1}^{M + 1} {\Pr ( {A_m}) }  = 1$ and $\Pr(\text{Outage})\le 1$.
\end{proof}
\vspace{-0mm}
\section{Extension of results to short-$L_\text{c}$ scenario}
In some parts of Sections III and V, the results were specialized for the long-$L_\text{c}$ scenario, i.e., they were obtained under the assumption that the channel does not change during a packet transmission period. This is an appropriate model for the slow-moving or stationary users where the channel experiences slow variations, e.g., \cite{5336856,wirelesskhodemun,Tcomkhodemun,1200407}. For the medium/fast speed users, on the other hand, the short-$L_\text{c}$ scenario is normally considered where the channel changes in each retransmission round independently \cite{tuninetti2011} or dependently \cite{ARQ20111,4200959,1379007,4356994}. In this case, since the length of the codewords are the same as the length of the fading block, the INR protocol is studied under fixed-length coding condition \cite{tuninetti2011,ARQ20111,4200959,1379007,4356994}. In the following, first the results of the RTD and INR protocols are restudied in the short-$L_\text{c}$ scenario. Then, Theorem 5 shows that the analytical arguments obtained for the long-$L_\text{c}$ scenario are also valid when the channel changes in each round.

Let the channel realization in the $n$-th (re)transmission round be $g_n$. Then, the received SNR at the end of the $m$-th RTD-based (re)transmission round is $\sum_{n=1}^{m}{g_n P_n}$ which is obtained by maximum ratio combining at the receiver. Therefore, while the probability terms $\Pr(\text{outage})^\text{RTD}$ and $\Pr(A_m)^\text{RTD}$, i.e., (11) and (13), are the same as before, the terms in (12) and (14) are respectively changed to
\vspace{-0mm}
\begin{align}
&\Pr {(m)^{\text{RTD}}} =\nonumber\\& \Pr \left(\log (1 + \sum_{n = 1}^{m - 1} {{g_n}{P_n}} ) < R \le \log (1 + \sum_{n = 1}^m {{g_n}{P_n}} )\right)
\end{align}
and
\vspace{-0mm}
\begin{align}
\Pr {(1,2,\ldots,m)^{\text{RTD}}} = \Pr \left(R \le \log (1 + \sum_{n = 1}^m {{g_n}{P_n}} )\right).
\end{align}
On the other hand, utilizing fixed-length coding INR protocol, (21) and (22) change to
\vspace{-0mm}
\begin{align}
&\Pr {(m)^{\text{INR}}} =\nonumber\\& \Pr \left(\sum_{n = 1}^{m - 1} {\log (1 + {g_n}{P_n}} ) < R \le \sum_{n = 1}^m {\log (1 + {g_n}{P_n}} )\right)
\end{align}
and
\vspace{-0mm}
\begin{align}
\Pr {(1,2,\ldots,m)^{\text{INR}}} = \Pr \left(R \le \sum_{n = 1}^m {\log (1 + {g_n}{P_n}} )\right),
\end{align}
respectively, if the channel changes in each (re)transmission round. In this way, the probability terms $\Pr(\text{outage})^\text{INR}$ and $\Pr(A_m)^\text{INR}$ are obtained with the same equations as before while (21) and (22) are replaced by (35) and (36), respectively\footnote{In this case, the probabilities $\Pr(\sum_{n=1}^{m}{\log(1+g_nP_n)}\le R)$ can be found by $m$-dimensional integration of the gain's pdf.}.


\emph{Theorem 5:} All assertions presented for the long-$L_\text{c}$ scenario are also valid in the short-$L_\text{c}$ scenario, i.e., when the channel changes in each retransmission round.
\begin{proof}
Please see the appendix.
\end{proof}
Note that, although the analytical results are same in these two scenarios, the numerical results are different for a given average power. However, as demonstrated in the following, the performance of the short- and long-$L_\text{c}$ scenarios can be mapped to each other if the average transmission power is scaled appropriately.

\emph{Theorem 6:} The performance of the considered ARQ protocols with uniform power allocation and short-$L_\text{c}$ fading model can be mapped to the one in a long-$L_\text{c}$ model with random power allocation and a different average transmission power.
\begin{proof}
Please see the appendix.
\end{proof}
In words, the theorem means that, although the channel remains fixed during a packet period of the long-$L_\text{c}$ scenario, we can use random power allocation to provide the same randomness as the one which is experienced in each (re)transmission round of the short-$L_\text{c}$ scenario. However, the average transmission power should be scaled appropriately. Finally, although the theorem is not proved for the cases with optimal power allocation (which is the main scope of the paper), it is interesting as it provides connection between the short- and long-$L_\text{c}$ scenarios.
\vspace{-0mm}
\section{Simulation results and discussions}
In this section, we illustrate the performance of the different studied cases. Since the number of studies is large, we have organized the text below into separate studies of RTD, of INR, of power allocation, of outage probability, and of different fading distributions; all these for varying feedback error conditions. In many cases, we have also investigated a wider range of parameters and fading conditions, but since the performance of those cases have followed the same trends as the ones shown, we have not included those results, to avoid unnecessary complexity.

As illustrated in \cite{641562,eurasipkhodemun}, Nakagami-$N$ distribution of the variable $g$ can model cases with different fading conditions where the fading severity decreases with $N$. For this reason, the simulation results of Figs. 1-9 are obtained for Nakagami-2 fading channel ${f_G}(g) = \frac{8}{{{w^2}}}{g^3}{e^{ - \frac{2}{w}{g^2}}},$ $g \ge 0,$ (moderate fading severity \cite{641562,eurasipkhodemun}) where we set $w=1$. Later, in Figs. 10-13, we study the effect of different fading distributions on the system performance. Moreover, in all figures except Figs. 2, 8 and 9, a maximum of $M=1$ bit feedback is considered.

\begin{figure}
\vspace{-0mm}
\centering
  \includegraphics[width=0.96\columnwidth]{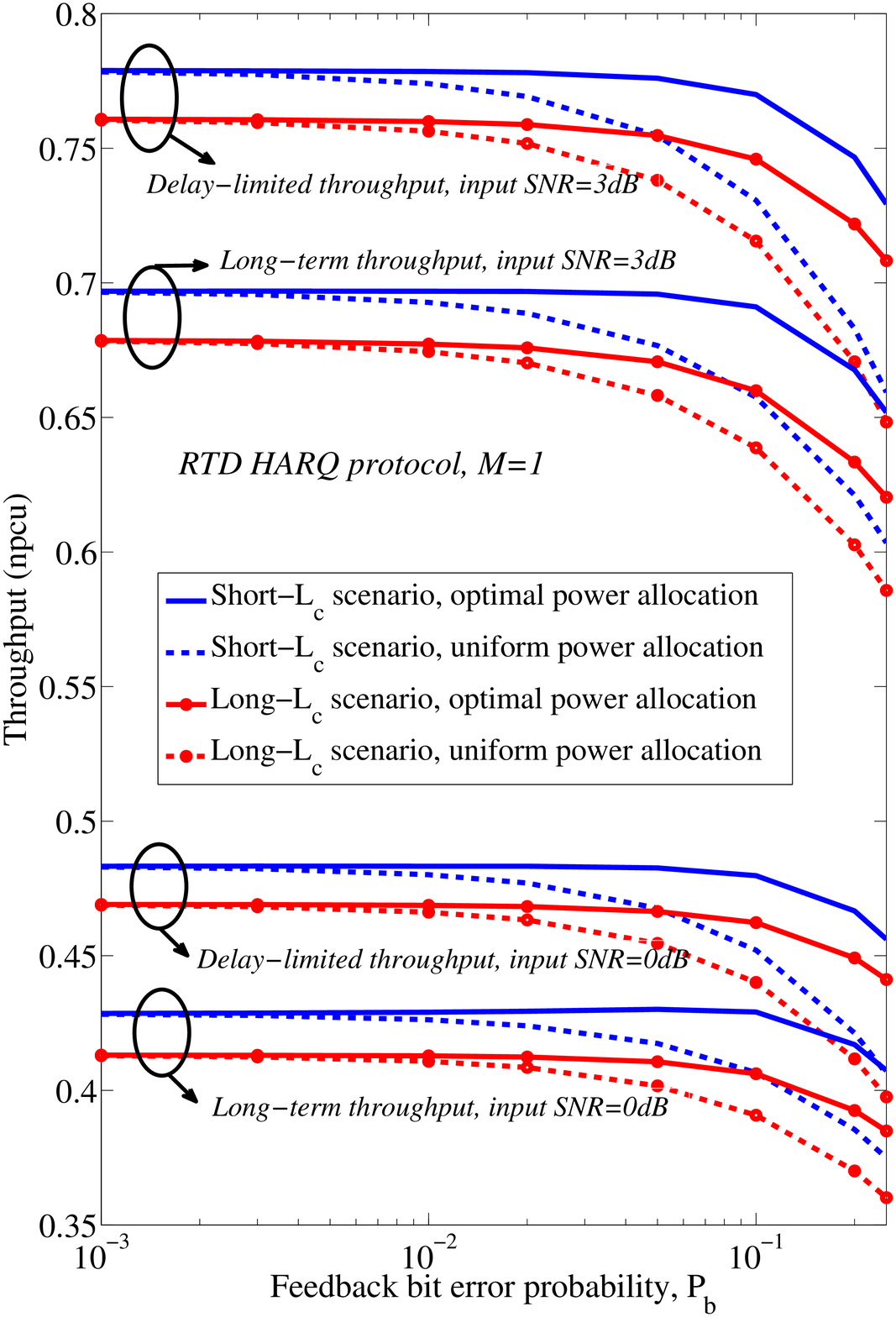}\\\vspace{-6mm}
\caption{Delay-limited and long-term throughput vs the feedback bit error probability $p_\text{b}$. RTD HARQ protocol, $M=1$. }\label{figure111}
\vspace{-2mm}
\end{figure}
\begin{figure}
\vspace{-0mm}
\centering
  \includegraphics[width=0.96\columnwidth]{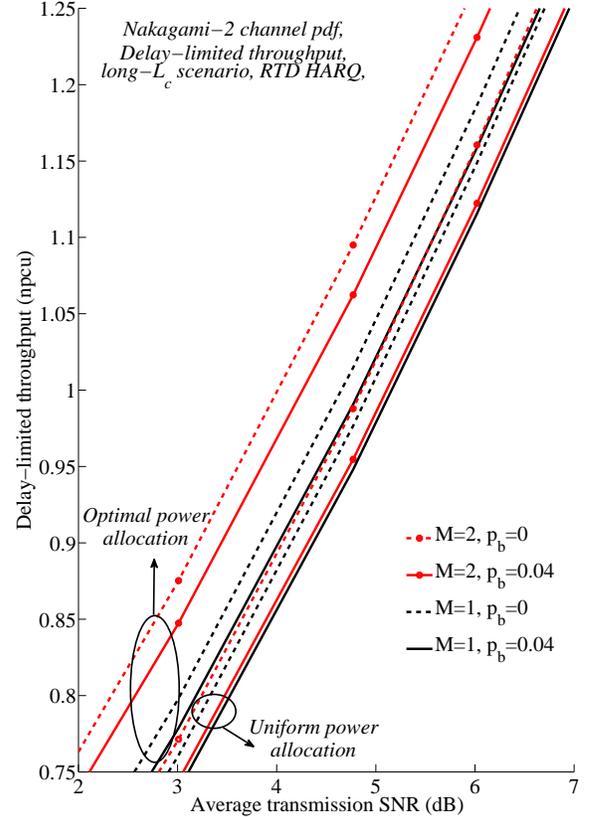}\\\vspace{-6mm}
\caption{Delay-limited throughput vs average transmission power. RTD HARQ protocol, $M=1$ and $2$, Nakagami-$2$ channel distribution, long-$\text{L}_\text{c}$ scenario. With optimal power allocation, considerable throughput increment is achieved by increasing the number of retransmissions.}\label{figure111}
\vspace{-2mm}
\end{figure}
\emph{Throughput in the RTD HARQ protocol:} Figs. 1 and 2 demonstrate the long-term and the delay-limited throughput of the noisy RTD HARQ protocol. According to the figures, the following points are concluded:
\begin{itemize}
  \item The robustness to feedback channel noise is slightly better in the long-$L_\text{c}$ scenario, compared to the short-$L_\text{c}$ model (Fig. 1).
  \item With optimal power allocation (resp. uniform power allocation), increasing the number of retransmissions leads to considerable (resp. marginal) throughput increase for both noisy and noise-free feedback conditions (Fig. 2). Also, the effect of optimal power allocation on the robustness of the system increases with $M$. Fig. 2 shows an example of this point where the throughput with $M=2$ and $p_\text{b}=0.04$ is less than the throughput achieved with $M=1$ and $p_\text{b}=0$, if uniform power allocation is considered. Finally, with Nakagami-2 fading channel, $M=1$ and in the practical range of feedback error probabilities, the throughput is not sensitive to optimal power allocation and the feedback bit error probability (Fig. 1).
\end{itemize}

\emph{Throughput in the INR HARQ protocol:} The effect of the feedback channel bit error probability on the long-term and delay-limited throughput of the INR scheme is studied in Figs. 3-5. Here, fixed-length coding is considered for the INR scheme, unless otherwise mentioned. The results show that:
\begin{itemize}
  \item
       With $M=1$ and uniform power allocation, the delay-limited throughput of the INR protocol decreases (almost) linearly with the feedback bit error probability $p_\text{b}$ (Fig. 3). However, the same as in the RTD, the throughput reduction due to erroneous feedback bits is negligible in different power allocation schemes, if $p_b$ is in the practical range of interest. However, it is later shown in Fig. 12 that, depending on the fading condition, there are cases where optimal power allocation can improve the \emph{relative} throughput, defined as the normalized difference between the throughput achieved by optimal and uniform power allocation.
  \item As expected, higher long-term and delay-limited throughput is achieved by the INR protocol, compared to the RTD (please see Section V and Fig. 4). However, the RTD protocol interestingly shows higher robustness to the feedback channel noise, specifically at high feedback bit error probabilities. This point can be seen in Fig. 4 where the throughput of the two schemes converge when the feedback bit error probability increases (The same points are valid for the short-$L_\text{c}$ scenario, although not seen in the figure.).
  \item The effect of variable-length coding on the performance of the INR scheme is demonstrated in Fig. 5. Compared to the fixed-length coding INR scheme, variable-length coding not only increases the throughput but also improves the robustness of the protocol to feedback bit errors.
\end{itemize}
\begin{figure}
\vspace{-0mm}
\centering
  \includegraphics[width=0.95\columnwidth]{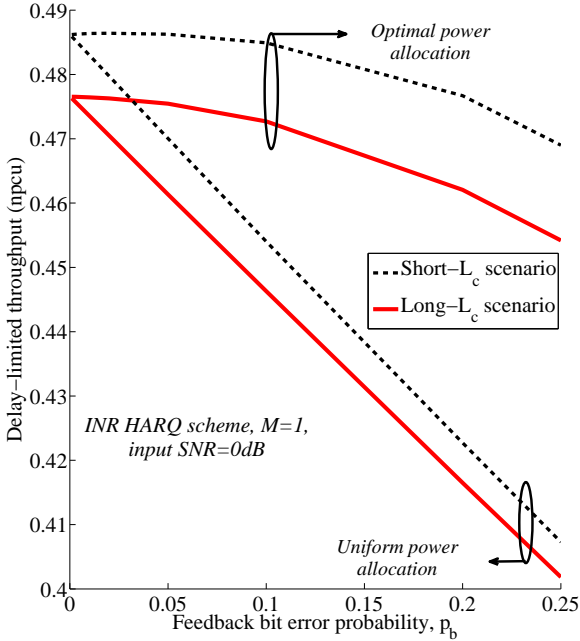}\\\vspace{-6mm}
\caption{Delay-limited throughput vs the feedback channel bit error probability $p_\text{b}$. Average transmission SNR $0dB$, INR HARQ protocol, $M=1$, Nakagami-$2$ channel model. }\label{figure111}
\vspace{-2mm}
\end{figure}
\begin{figure}
\vspace{-0mm}
\centering
  \includegraphics[width=0.95\columnwidth]{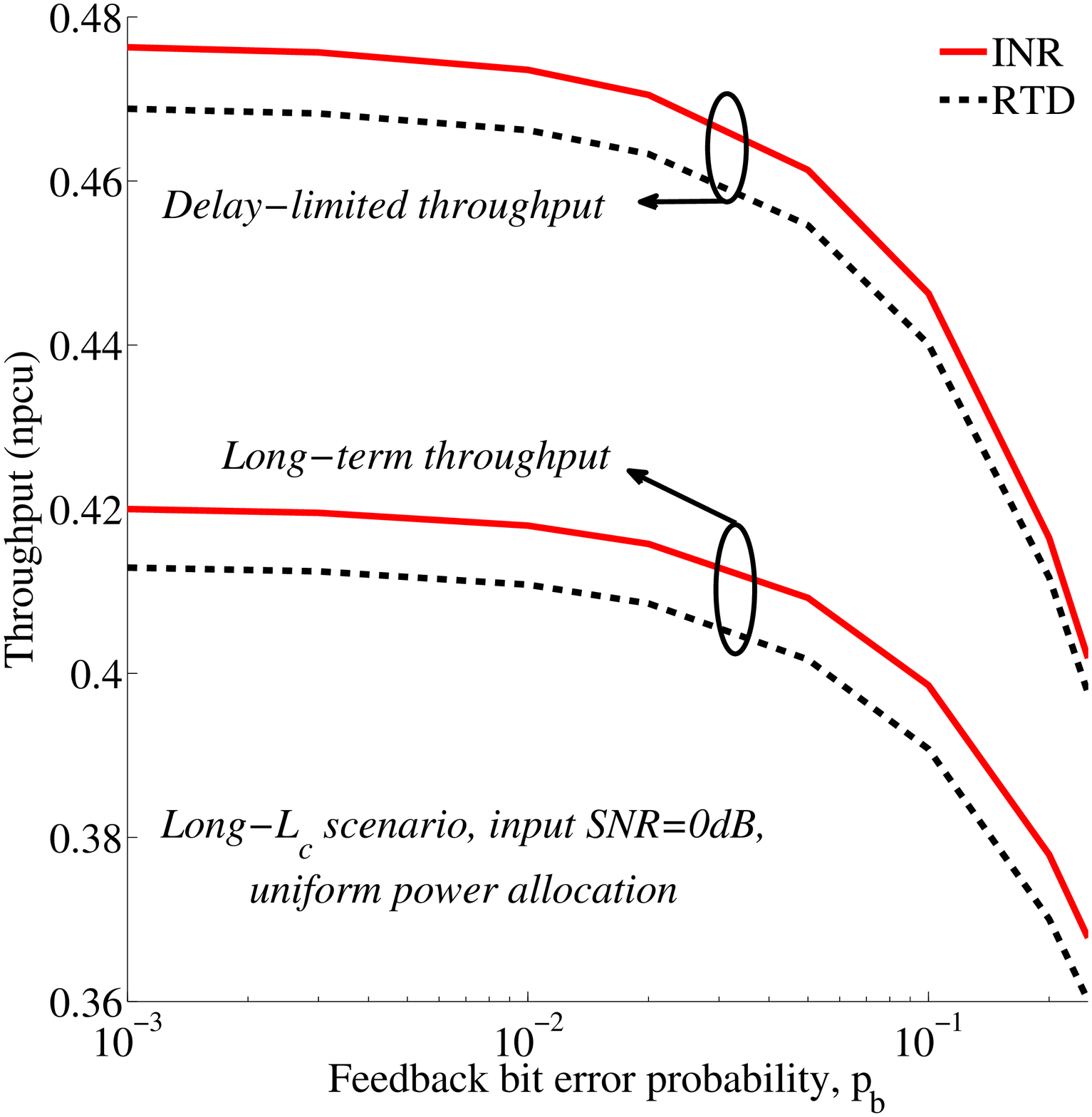}\\\vspace{-5mm}
\caption{Delay-limited and long-term throughput vs the feedback bit error probability $p_\text{b}$. Long-$L_\text{c}$ scenario, uniform power allocation. Higher throughput is achieved by the INR scheme, compared to the RTD. With high feedback bit error probability, however, the difference between the throughput of the two HARQ schemes decreases. }\label{figure111}
\vspace{-3mm}
\end{figure}
\begin{figure}
\vspace{-0mm}
\centering
  \includegraphics[width=0.95\columnwidth]{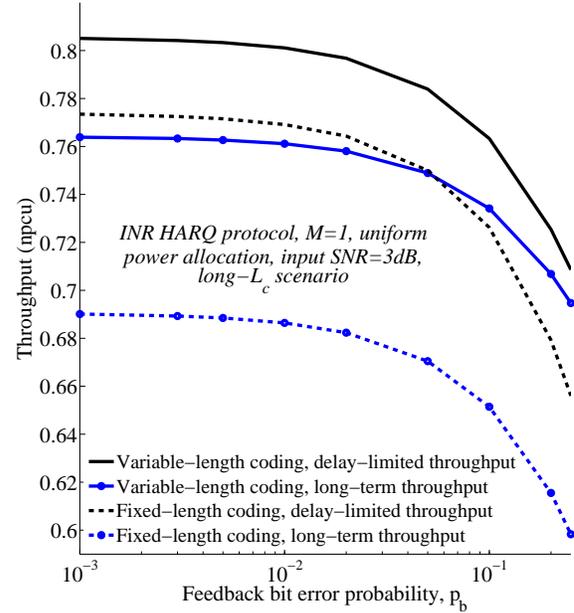}\\\vspace{-6mm}
\caption{The effect of variable- and fixed-length coding on the throughput of the INR HARQ protocol in noisy feedback condition. Uniform power allocation, $3dB$, a maximum of $M=1$ retransmission round, long-$L_\text{c}$ scenario. Variable-length coding increases the throughput and the robustness of the INR protocol to feedback bit errors. }\label{figure111}
\vspace{-2mm}
\end{figure}

\emph{On the optimal (re)transmission powers:} Figs. 6-8 show the optimal (re)transmission powers maximizing the throughput of different HARQ protocols. Here, the following points are deduced from the figures\footnote{Although Figs. 6 and 8 do not include the results for the long-term throughput (and Fig. 7 does not demonstrate the optimal powers for the RTD protocol), the simulations show the same qualitative conclusions for the not-included cases.}:
\begin{itemize}
  \item Compared to the short-$L_\text{c}$ scenario, the difference between the (re)transmission powers of different rounds is higher in the long-$L_\text{c}$ scenario (Figs. 6 and 7).
  \item The difference between the (re)transmission powers of different rounds increases with the feedback channel bit error probability (Fig. 7). Also, the power allocated to the latest retransmission rounds decreases when the feedback channel becomes noisy. Intuitively, this is because with worse feedback channel condition it is tried to decode the data in the first round(s), so that the dependency to the feedback signal is reduced (Figs. 6-8).
  \item For both INR and RTD and $M=1$ and $2$ cases, the power terms are observed to increase with the average transmission power (almost) linearly (Figs. 6-8). However, there is no general relationship between the optimal power terms (Also see \cite{5771499,ARQGlarsson} for further discussion about the unexpected behavior of optimal power terms in noise-free feedback condition.).
\end{itemize}

\begin{figure}
\vspace{-0mm}
\centering
  \includegraphics[width=0.95\columnwidth]{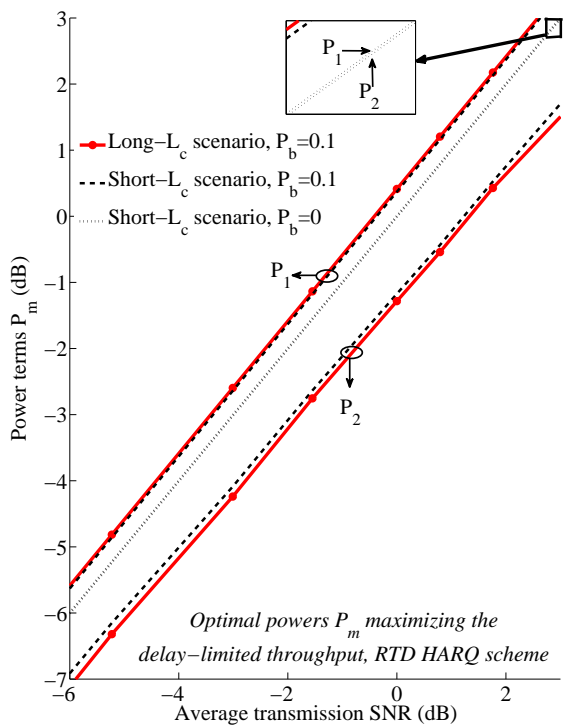}\\\vspace{-5mm}
\caption{Optimal transmission powers maximizing the delay-limited throughput in the RTD protocol, $M=1$, Nakagami-$2$ channel model. The same trend is observed when maximizing the long-term throughput. The difference between the transmission powers increases with the feedback channel bit error probability.  }\label{figure111}
\vspace{-2mm}
\end{figure}
\begin{figure}
\vspace{-0mm}
\centering
  \includegraphics[width=0.95\columnwidth]{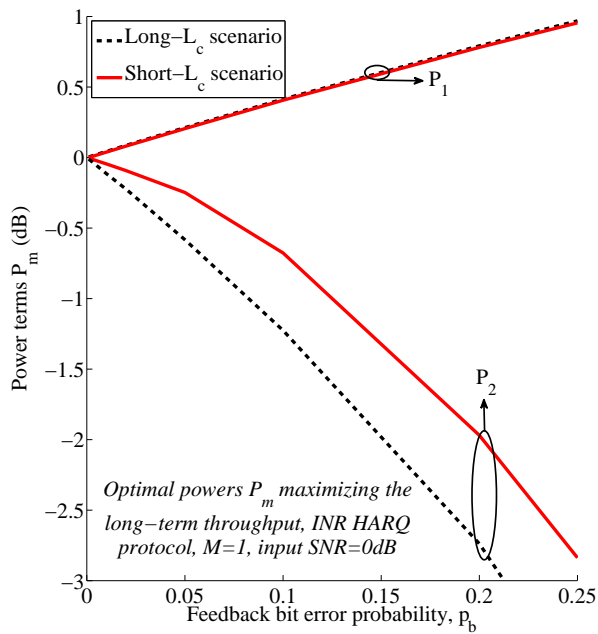}\\\vspace{-5mm}
\caption{Optimal (re)transmission powers, maximizing the long-term throughput, vs the feedback bit error probability $p_\text{b}$. Average transmission SNR $0dB$, INR HARQ protocol, $M=1$, Nakagami-$2$ channel model. The more noisy the feedback channel is, the more difference is observed between the powers of the two rounds. }\label{figure111}
\vspace{-2mm}
\end{figure}
\begin{figure}
\vspace{-0mm}
\centering
  \includegraphics[width=0.95\columnwidth]{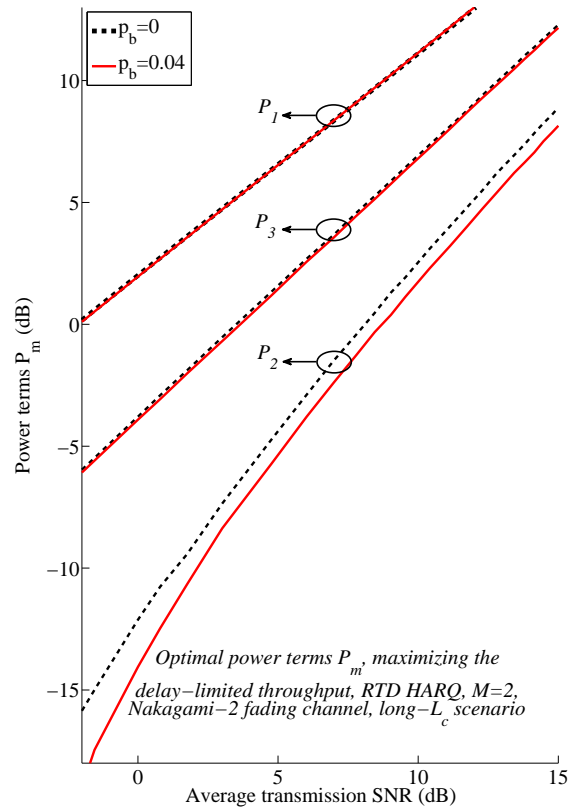}\\\vspace{-5mm}
\caption{Optimal transmission powers, maximizing the delay-limited throughput. RTD HARQ protocol, $M=2$, long-$L_\text{c}$ scenario, Nakagami-$2$ channel model. }\label{figure111}
\vspace{-2mm}
\end{figure}

\emph{Outage probability with a noisy HARQ protocol:} Figs. 9 and 10 demonstrate the system outage probability in the presence of noisy RTD and INR HARQ, respectively.
Here, the results are obtained for a fixed initial transmission rate $R=0.4$. The figures emphasize the following points:
\begin{itemize}
  \item Feedback channel bit error probability leads to considerable outage probability increment, particularly when $M$ increases (Fig. 9). The sensitivity of the outage probability to the feedback channel noise is intuitively due to the fact that the outage probability is determined by the small portion of the packets which can not be decoded correctly. Therefore, even a few numbers of failures, which may occur due to erroneous feedback, become important and affect the outage probability considerably. Thus,  although adaptive power allocation reduces the effect of feedback channel noise, the outage probability is still sensitive to $p_b.$
  \item With a noise-free feedback channel, power allocation is very effective in outage probability reduction of the HARQ protocols, specifically when the transmission power increases (Figs. 9 and 10. See also \cite{5771499,ARQGlarsson}.). On the other hand, for noisy feedback channel the reduction of the outage probability is less pronounced. However, optimal power allocation is still very useful, particularly at high SNR.
\end{itemize}
\begin{figure}
\vspace{-0mm}
\centering
  \includegraphics[width=0.96\columnwidth]{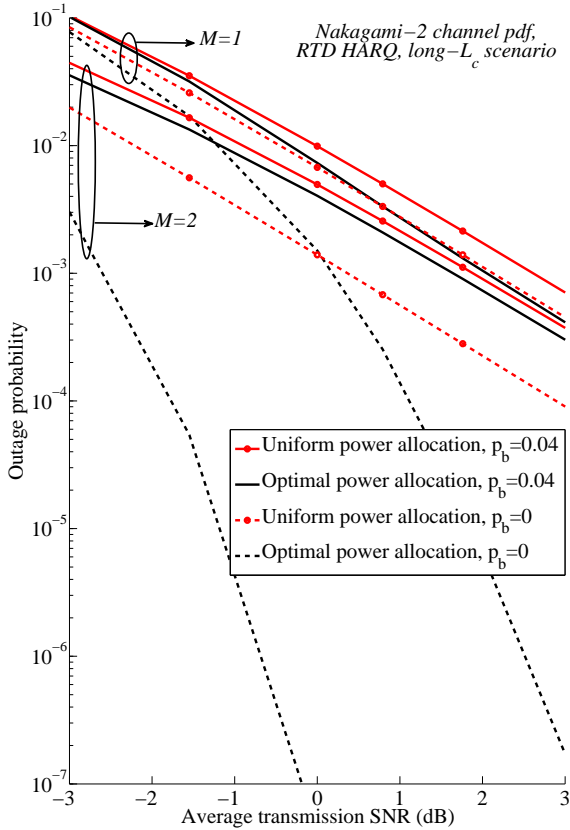}\\\vspace{-6mm}
\caption{Outage probability vs the average transmission SNR. RTD HARQ protocol, $M=1$ and $2$, $R=0.4$, Nakagami-$2$ channel model, long-$L_\text{c}$ scenario. With a noise-free feedback channel, optimal power allocation leads to substantial outage probability reduction, particularly when the transmission power increases. However, compared to noise-free condition, the effect of power allocation on the outage probability of the RTD scheme decreases in noisy feedback conditions.}\label{figure111}
\vspace{-2mm}
\end{figure}
\begin{figure}
\vspace{-0mm}
\centering
  \includegraphics[width=0.96\columnwidth]{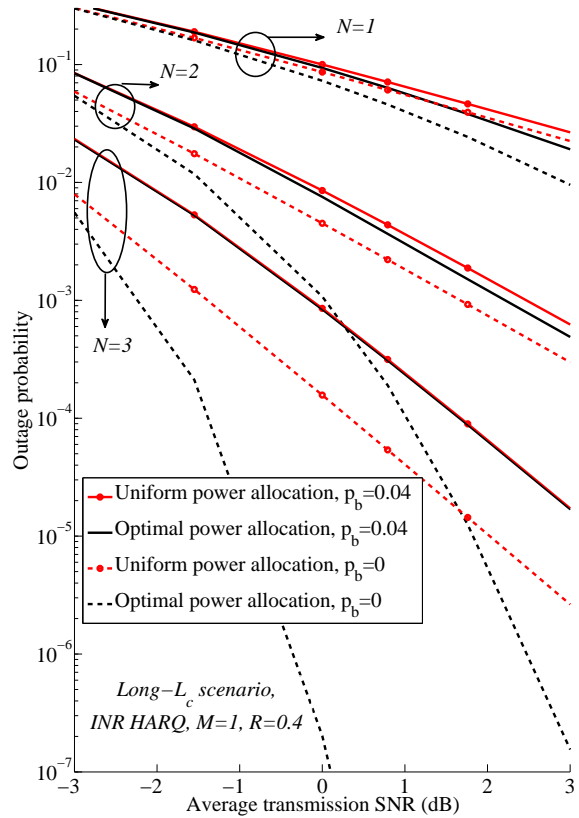}\\\vspace{-6mm}
\caption{Outage probability vs the average transmission SNR. INR HARQ protocol, $M=1$, $R=0.4$ (fixed-length coding) and different Nakagami-$N$ channel models. For all considered channel models, optimal power allocation results in substantial outage probability reduction, if the feedback bits are received error-free. However, the effect of power allocation on the outage probability of the INR scheme decreases in noisy feedback conditions, particularly when $N$ increases.}\label{figure111}
\vspace{-2mm}
\end{figure}

\emph{On the effect of fading distribution:} As illustrated throughout the paper, the data transmission efficiency of HARQ protocols depends on the fading pdf. For this reason, in Figs. 10-13 we study the performance of HARQ schemes in different Nakagami-$N$ channel models. In Fig. 11, we present a $\beta$\emph{-region} which is defined as
\vspace{-0mm}
\begin{align}
&\varphi(\beta)\doteq \bigg\{(p_\text{b},\text{SNR}): \nonumber\\&\,\,\,\,\,\,\,\,\,\,\,\, \frac{\Pr(\text{Outage}|p_\text{b},\text{SNR})-\Pr(\text{Outage}|p_\text{b}=0,\text{SNR})}{\Pr(\text{Outage}|p_\text{b}=0,\text{SNR})}\le \beta\%\bigg\}.\nonumber
\end{align}
Each curve in Fig. 11 defines the set of feedback bit error probabilities (for a given SNR) which allow performance within $(1-\beta)\%$ of the performance of the noise-free system. This set of probabilities correspond to the area below each curve. We set $\beta=5\%$ in the figure. Moreover, Fig. 12 shows the relative throughput difference
\vspace{-0mm}
 \begin{align}
\zeta=\frac{\eta_\text{LT}-\eta_{\text{LT},p_b=0, P_m=P,\forall m}}{\eta_{\text{LT},p_b=0, P_m=P,\forall m}}\%\nonumber
\end{align}
where $\eta_{\text{LT},p_b=0, P_m=P,\forall m}$ denotes the throughput that is achieved with uniform power allocation and a noise-free feedback channel. In other words, Fig. 12 demonstrates the relative gain of optimal power allocation in terms of throughput. Finally, setting $R=1$, Fig. 13 demonstrates the \emph{usefulness} region of the noisy ARQ protocols, in terms of throughput, in comparison to the open-loop communication setup. That is, the area below each curve corresponds to the set of pairs $(p_b,\text{SNR})$ for which the noisy ARQ leads to higher throughput, compared to an open-loop communication scheme. The following points are deduced from the figures:
\begin{itemize}
  \item The general behavior of the noisy and noise-free HARQ protocols is the same in different fading channels. For instance, optimal power allocation leads to considerable outage probability reduction in different channel models, while the effect of power allocation decreases at high $p_b$'s. (Fig. 10).
  \item For different Nakagami-$N$ channels, the sensitivity to feedback channel noise increases with the transmission power, if uniform power allocation is considered (Fig. 11). However, even with high $p_\text{b}$'s, it is still possible to reach a large portion of the noise-free system performance if the (re)transmission powers are selected optimally (Figs. 11 and 12).
  \item Compared to open-loop communication, the optimal power allocation increases the robustness of the noisy ARQ schemes and the sensitivity of the throughput to feedback channel noise increases with the SNR (Fig. 13).
\end{itemize}
\begin{figure}
\vspace{-0mm}
\centering
  \includegraphics[width=0.96\columnwidth]{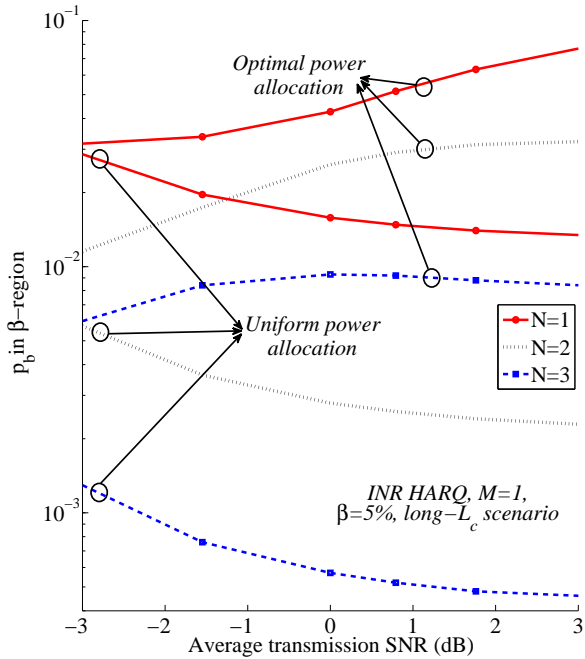}\\\vspace{-5mm}
\caption{$\beta$-region for the outage probability of the INR HARQ protocol and different Nakagami-$N$ channel models. $M=1$, $R=0.4$, long-$L_\text{c}$ scenario. Below each curve gives the feedback bit error probabilities below which $(1-\beta)\%$ of the noise-free channel performance is achievable ($\beta=5\%$). }\label{figure111}
\vspace{-2mm}
\end{figure}
\begin{figure}
\vspace{-0mm}
\centering
  \includegraphics[width=0.96\columnwidth]{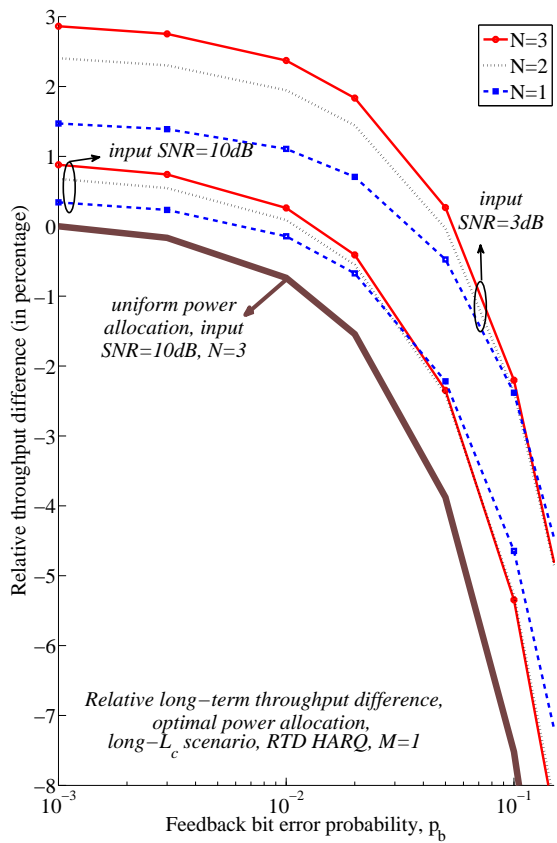}\\\vspace{-5mm}
\caption{Relative throughput difference vs the feedback channel bit error probability. RTD HARQ, $M=1$, long-$L_\text{c}$ scenario, different Nakagami-$N$ channel models, optimal power allocation. }\label{figure111}
\vspace{-2mm}
\end{figure}
\begin{figure}
\vspace{-0mm}
\centering
  \includegraphics[width=0.96\columnwidth]{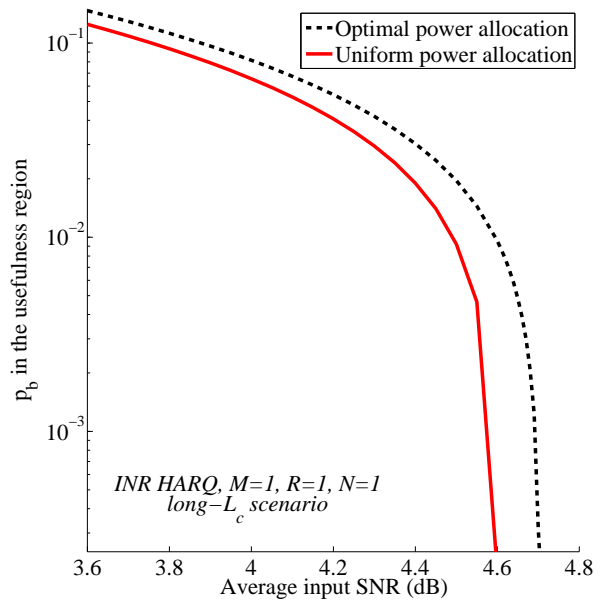}\\\vspace{-5mm}
\caption{Usefulness region. The area below each curve corresponds to the set of pairs $(p_b,\text{SNR})$ for which the noisy ARQ leads to higher throughput, compared to an open-loop communication scheme. INR HARQ, $M=1$, long-$L_\text{c}$ scenario, Nakagami-$1$ channel model. The results are for a fixed (non-optimized) initial transmission rate $R=1$.}\label{figure111}
\vspace{-0mm}
\end{figure}

Finally, considering all simulation results, the following points are interesting to be noted: 1) compared to the long-$L_\text{c}$ scenario, lower outage probability and higher throughput is achieved by the HARQ protocols in the short-$L_\text{c}$ scenario. This is intuitively due to the fact that more time diversity is exploited by the HARQ schemes when the channel changes in the retransmission rounds. However, the qualitative conclusions drawn from the simulations are valid for both scenarios. 2) The delay-limited throughput exceeds the long-term throughput in all conditions (See also Theorem 4).
\vspace{-0mm}
\section{Conclusion}
Considering different power allocation capabilities, this paper studied the performance of the ARQ protocols in noisy feedback conditions. The results indicate that the ARQ protocols are not very sensitive to optimal power allocation and feedback channel noise, when the goal is to maximize the throughput in a practical range of feedback bit error probabilities. However, the effect of optimal power allocation on the throughput and system robustness increases with the number of retransmissions, and the efficiency of the power allocation schemes depends on the fading condition. Optimal power allocation plays an important role on the outage probability of noisy ARQ protocols, although the effect of optimal power allocation decreases at high feedback error probabilities.
Also, the difference between the optimal (re)transmission powers increases with the feedback bit error probability and decreases with the forward channel variability. Many analytical assertions about the ARQ protocols are valid independent of the forward channel characteristics. Finally, although the INR HARQ outperforms the RTD method in many aspects, the difference between these methods decreases when the transmission power decreases or the feedback bit error probability increases.
\vspace{-0mm}
\section{appendices}
\vspace{-0mm}
\subsection{Performance analysis for basic ARQ protocols}

In basic ARQ protocols with adaptive power allocation, the transmitter keeps sending scaled versions of the same codeword in the (re)transmission rounds and the receiver decodes only the most recently received signal, regardless of the previously received signals.

For a noise-free feedback channel and in short-$L_\text{c}$ scenario, \cite{4200959} has previously shown that the transmission powers in the basic ARQ scheme should increase with the number of retransmission round if the goal is to minimize the outage probability. Moreover, as mentioned in \cite{5336856}, considering the long-$L_\text{c}$ scenario there is no use in basic ARQ if uniform power allocation is implemented. The following lemma shows that, independent of the optimization objective function and for any feedback channel conditions, the transmission powers in the basic ARQ scheme must be increasing in every round, if the channel remains fixed within all (re)transmissions.

\emph{Lemma 2:} Consider the long-$L_\text{c}$ scenario. In basic ARQ schemes the transmission powers must be increasing in every retransmission, independent of the optimization objective function.

\begin{proof}
Using basic ARQ, the data is decodable at the $m$-th (re)transmission round if $\log(1+gP_m)\ge R$ where $R=\frac{Q}{L}$ is the initial codeword rate. Therefore, given that the codeword is not decodable at the $m$-th round, i.e., $g< \frac{e^R-1}{P_m}$, retransmitting it with lower (or equal) power at the $(m+1)$-th round is useless as $\Pr(g< \frac{e^R-1}{P_{m+1}} |g< \frac{e^R-1}{P_m} \,\&\, P_m>P_{m+1})=1$. Therefore, to have some chance for decoding the data, we should have $P_m\le P_{m+1},\, \forall m$. Finally, note that the lemma is valid independent of the optimization criterion.
\end{proof}


Using Lemma 2, the long-term throughput and the average transmission power for the basic ARQ protocol in long-$L_\text{c}$ scenario are obtained with the same equations as for the RTD scheme, i.e., (9)-(11) and (13), while the probability terms (12) and (14) are respectively replaced by
\vspace{-0mm}
\begin{align}
\Pr {( m) ^{\text{Basic}}} &= \Pr \left( \log (1 + g{P_{m - 1}}) < R \le \log (1 + g{P_m})\right) \nonumber\\& = {F_G}(\frac{{{e^R} - 1}}{{{P_{m - 1}}}}) - {F_G}(\frac{{{e^R} - 1}}{{{P_m}}})
\end{align}
and
\vspace{-0mm}
\begin{align}
\Pr {( 1,\ldots,m) ^{\text{Basic}}} &= \Pr ( R \le \log (1 + g{P_m}))  \nonumber\\&= 1 - {F_G}(\frac{{{e^R} - 1}}{{{P_m}}}).
\end{align}
On the other hand, assuming short-$L_\text{c}$ scenario, (37) and (38) are respectively rephrased as
\vspace{-0mm}
\begin{align}
&\Pr (m)^{\text{Basic}} \nonumber\\&= \Pr \big(\log (1 + {g_n}{P_n}) < R\,\forall n < m\,\& \,\log (1 + {g_m}{P_m}) \ge R\big) \nonumber\\&= \prod_{n = 1}^{m - 1} {{F_G}(\frac{{{e^R} - 1}}{{{P_n}}})} \left(1 - {F_G}(\frac{{{e^R} - 1}}{{{P_m}}})\right),
\end{align}
\vspace{-0mm}
\begin{align}
\Pr (1,\ldots,m)^{\text{Basic}} &= \Pr \left(\exists n \le m,\,\log (1 + {g_n}{P_n}) \ge R\right) \nonumber\\&= 1 - \prod_{n = 1}^m {{F_G}(\frac{{{e^R} - 1}}{{{P_n}}})}.
\end{align}
\vspace{-0mm}
\subsection{Proof of Theorem 1}
If the data (re)transmission ends at the $m$-th, $m \le M$, (re)transmission round, $m$ feedback bits are sent to the transmitter. Also, $M$ bits are fed back if all the $M+1$ possible (re)transmission rounds are used. Therefore, the feedback load is
\vspace{-0mm}
\begin{align}
\bar B = \sum_{m = 1}^M {m\Pr ( {A_m}) }  + M\Pr ( {A_{M + 1}}).
\end{align}
Moreover, the expected number of (re)transmission rounds is found as $\bar r = \sum_{m = 1}^{M + 1} {m\Pr ( {A_m}) }$.\footnote{Note that $\bar r=\bar B+\Pr(A_{M+1})$. That is, the only difference between the expected number of retransmission rounds and the feedback load is in the last (re)transmission round where, while the data is retransmitted, no feedback is sent to the transmitter.}

Now, assume that setting $P_m=P\, \forall m$ the optimal initial transmission rate of the RTD scheme maximizing the throughput has been obtained to be $R=\hat R$. We set the equivalent transmission rates of the INR scheme such that $\frac{{{e^{{R^{(m)}}}} - 1}}{P} = \frac{{{e^{\hat R}} - 1}}{{mP}}$, which is not necessarily optimal for the INR. In this case, from (15), (16), (19) and (20), the probability terms $\Pr(\text{Outage})$ and $\Pr(A_m)$ and, consequently, the feedback load and the the expected number of (re)transmission rounds of both schemes are the same. Therefore, Theorem 1 is proved if we show that in each (re)transmission round the equivalent transmission rate in the INR protocol is higher than the equivalent rate in the RTD scheme, i.e., ${R^{(m)}} = \log (1 + \frac{{{e^{\hat R}} - 1}}{m}) \ge \frac{\hat R}{m}$. To show this, we define $\omega(R) = \log (1 + \frac{{{e^R} - 1}}{m}) - \frac{R}{m}$ where $\omega(0) = 0$ and $\frac{{\partial \omega}}{{\partial R}} \ge 0$. Therefore, $\omega(R)\ge 0\,\forall R\ge 0$, i.e., ${R^{(m)}} = \log (1 + \frac{{{e^{\hat R}} - 1}}{m}) \ge \frac{\hat R}{m}$. Consequently, using (4) the throughput of the INR scheme is higher than the throughput of the RTD, as the denominator of (4) is smaller for the INR.
\vspace{-2mm}
\subsection{Proof of Theorem 2}
For simplicity, we prove the theorem assuming fixed-length coding for the INR protocol. Then, as fixed-length coding is a special case of the general INR protocol, the theorem is also proved for the general case. Assume that the transmission parameters $R$, $P_m,\,m=1,\ldots,M+1$ have been optimized in terms of the RTD-based power-limited throughput, i.e., when RTD is considered in (9) and (10). We consider the same transmission rate and powers for fixed-length INR, which is not necessarily optimal for this protocol.

Given that the data (re)transmission is stopped at the of the $m$-th round, the transmission energy is $\xi^{(m)}=L\sum_{n=1}^{m}{P_n}$ for both protocols. On the other hand, we have $\Pr(\text{successful decoding}|A_m)=1-\Pr(\text{failure}|A_m)$ and
\vspace{-0mm}
\begin{align}
\Pr(\text{failure}|A_m)=\left\{\begin{matrix}
\Pr(\log(1+g\sum_{n=1}^{m}{P_n})<R) & \text{For RTD}\\
\Pr(\sum_{n=1}^{m}{\log(1+gP_n)}<R) & \text{For INR}
\end{matrix}\right.
\end{align}
where, from (23), we have $\Pr(\text{failure}|A_m)^\text{INR}\le\Pr(\text{failure}|A_m)^\text{RTD}$. Thus, with the same channel conditions, transmission parameters and number of channel uses, it is always more likely to successfully decode the data by the INR which leads to higher throughput in the INR, when compared with the RTD. Also, the average transmission power of the INR does not exceed the one in the RTD because, in the worst case, we can continue the data (re)transmission with the same procedure as in the RTD.

Finally, it is worth noting that the superiority of the INR over RTD is due to the fact that a \emph{better} code is implemented in the INR, compared to the RTD. Therefore, we can use the same arguments as in the theorem to show that the INR outperforms the RTD in terms of different metrics, which is of course at the cost of the encoding/decoding complexity (Please see \cite{tuninetti2011,5336856,Tcomkhodemun} for further comparisons between the RTD and INR.).

\vspace{-0mm}
\subsection{Proof of Theorem 5}

The difference between the short- and long-$L_\text{c}$ scenarios is in the calculation of the probabilities $\Pr {(m)}$ and $\Pr {(1,2,\ldots,m)}$. Therefore, the proofs of Lemma 1, Corollaries 1-3 and Theorem 4 part (I) are applicable in the short-$L_\text{c}$ scenario as well, since the arguments are valid for every given probability terms $\Pr(A_m)$ and $\Pr(S_m)$, independent of how they are found. Also, Theorem 3 can be proved with exactly the same procedure as for the long-$L_\text{c}$ scenario, since the approximations (24) and (25) can be applied in (33)-(36) as well.

Equation (23) can be used to prove Theorems 1, 2 and 4 under the condition that the channel changes in each retransmission round; Implementing (33)-(36) in the proof of Theorem 2, the terms $\Pr(\text{failure}|A_m)^{\text{RTD}}$ and $\Pr(\text{failure}|A_m)^{\text{INR}}$ are replaced by ${\gamma ^{\text{RTD}}} = \Pr(\log(1+\sum_{n=1}^{m}{g_nP_n})<R)$ and ${\gamma ^{\text{INR}}} = \Pr(\sum_{n=1}^{m}{\log(1+g_nP_n)}<R)$, respectively, where according to (23) we have ${\gamma ^{\text{INR}}} \le {\gamma ^{\text{RTD}}}$. This is the only modification required for the short-$L_\text{c}$ scenario and the rest of the proof does not need to be changed.


Theorem 1 was previously proved using variable-length coding in the INR protocol. In order to prove it in the short-$L_\text{c}$ scenario, where variable-length coding is not realistic, the proof is changed as follows. Let $p_\text{b} \le \frac{1}{2}$, otherwise the feedback bits are reversed. With some manipulations, the outage probability with uniform power allocation is obtained as
\vspace{-0mm}
\begin{align}
\Pr ( \text{Outage})  = {p_\text{b}}\sum_{m = 1}^M {{{(1 - {p_\text{b}})}^{m - 1}}{\alpha _m}}  + {(1 - {p_\text{b}})^M}{\alpha _{M + 1}}
\end{align}
where, considering the short-$L_\text{c}$ scenario, we have $\alpha _m^{\text{RTD}} = \Pr \{ \log (1 + \sum_{n = 1}^m {{g_n}P} ) \le R\}$ and $\alpha _m^{\text{INR}} = \Pr \{ \sum_{n = 1}^m {\log (1 + {g_n}P)}  \le R\} $. Thus, according to (23), we have $\alpha _m^{\text{INR}} \le \alpha _m^{\text{RTD}}$,
%
%
that is, $\Pr {(\text{outage})^{\text{INR}}} \le \Pr {(\text{outage})^{\text{RTD}}}$ for a given rate $R$. Therefore, from (9), Theorem 1 is proved under the short-$L_\text{c}$ assumption if ${C^{\text{INR}}} = \sum_{m = 1}^{M + 1} {m\Pr {{({A_m})}^{\text{INR}}}}  \le \sum_{m = 1}^{M + 1} {m\Pr {{({A_m})}^{\text{RTD}}}}  = {C^{\text{RTD}}}$, i.e., we have to show that the denominator of the throughput function in (9) is less for the INR when compared with the RTD. However, considering $\alpha _m^{\text{RTD}} = \Pr \{ \log (1 + \sum_{n = 1}^m {{g_n}P} ) \le R\}$ and $\alpha _m^{\text{INR}} = \Pr \{ \sum_{n = 1}^m {\log (1 + {g_n}P)}  \le R\} $, it can be written
\vspace{-0mm}
\begin{align}
{C^\text{H}} &= \sum_{m = 1}^{M + 1} {m\Pr {{({A_m})}^\text{H}}}  = 1 + \sum_{m = 1}^M {\Pr {{({{\bar A}_1},\ldots,{{\bar A}_m})}^\text{H}}} \nonumber\\&\mathop  = \limits^{(e)} \sum_{j = 0}^M {p_b^j}  + (1 - 2{p_b})\sum_{m = 1}^M {{{(1 - {p_\text{b}})}^{m - 1}}(\sum_{j = 0}^{M - m} {p_\text{b}^j} )\alpha _m^\text{H}},
\end{align}
where $\text{H}=\{\text{RTD},\text{INR}\}$. Also, $\Pr {{({{\bar A}_1},\ldots,{{\bar A}_m})}^\text{H}}$ is the probability that, implementing $\text{H}=\{\text{RTD},\text{INR}\}$ HARQ protocol, the data (re)transmission does not stop in the $n=1,\ldots,m$ rounds. Then, $(e)$ follows from the fact that using, e.g., (13), (33) and (34), the probability term $\Pr {{({{\bar A}_1},\ldots,{{\bar A}_m})}^\text{H}}$ is found as
\vspace{-0mm}
\begin{align}
\Pr {({{\bar A}_1},\ldots,{{\bar A}_m})^\text{H}} &= \sum_{n = 1}^m {\Pr (n){{(1 - {p_\text{b}})}^{n - 1}}p_\text{b}^{m + 1 - n}}  \nonumber\\&+ (1 - \Pr (1,\ldots,m)){(1 - {p_\text{b}})^m} \nonumber\\&= p_\text{b}^m + (1 - 2{p_\text{b}})\sum_{n = 1}^m {\alpha _m^\text{H} p_\text{b}^{m - n}{{(1 - {p_\text{b}})}^{n - 1}}}.
\end{align}
Note that, according to (44), $C^\text{H}$ is an increasing function of $\alpha_m^\text{H}$. Therefore, since $\alpha _m^{\text{INR}} \le \alpha _m^{\text{RTD}}$, we have $C^\text{INR}\le C^\text{RTD}$. Hence, with the same initial codeword rate $R$, the INR protocol outperforms the RTD in terms of throughput and feedback load, as not only less outage probability is achieved by the INR but also it leads to less expected number of (re)transmission rounds when compared with RTD. Finally, the same procedure as in (44) can be implemented in the $\Pr(S_m)$ to prove Theorem 4 part (II) under short-$L_\text{c}$ assumption.
\vspace{-0mm}
\subsection{Proof of Theorem 6}
For simplicity, we prove the theorem for the INR ARQ protocol; as illustrated throughout the paper, the only difference between the two considered cases is in the probability terms
\vspace{-0mm}
\begin{align}\label{eq:eqproof4}
\alpha_m^{\text{short}-L_\text{c}, \text{uniform power}}=\Pr\{\sum_{n=1}^{m}{\log(1+g_nP)}\le R\}
\end{align}
\vspace{-0mm}
\begin{align}\label{eq:eqproof42}
\alpha_m^{\text{long}-L_\text{c}, \text{random power}}=\Pr\{\sum_{n=1}^{m}{\log(1+g\tilde P_n)}\le R\}.
\end{align}
Thus, the performance, e.g., throughput and the outage probability, of the two cases is the same if the random powers ($\tilde P_n$ in (47)) are selected via a specific distribution $F_\varepsilon$ such that $F_{1+g_nP}(x)=F_{1+g\tilde P_n}(x),\forall x,$ i.e., the same randomness is experienced in the channel quality of the two cases. Here, the only point is that the average power in the second case is
\vspace{-0mm}
\begin{align}\label{eq:averagepowerrandom}
\Phi^{\text{long}-L_\text{c}, \text{random power}}=E_\varepsilon\{\frac{{\sum_{m = 1}^{M + 1} {{\tilde P_m}\left(1 - \sum_{n = 1}^{m - 1} {\Pr ( {A_n}) } \right)} }}{{\sum_{m = 1}^{M + 1} {m\Pr ( {A_m})} }}\}
\end{align}
which is different from the transmission power in the first case, i.e., $\Phi^{\text{short}-L_\text{c}, \text{uniform power}}=P$. Finally, note that (48) comes from (5) with fixed-length coding and the expectation is on $F_\varepsilon$.
\vspace{-0mm}
\bibliographystyle{IEEEtran} 
\bibliography{masterISITrevised}

\vfill

\end{document}